\documentclass[manuscript]{aastex631}

\usepackage{amsmath}

\graphicspath{{./}{figures/}}

\begin{document}

\title{Modeling the UV/optical light curve of re-brightening tidal disruption events}

\author[0000-0003-4121-5684]{Shiyan Zhong}
\affiliation{South-Western Institute for Astronomy Research, Yunnan University, Kunming, 650500 Yunnan, China}

\begin{abstract}
In recent years, a new subclass of tidal disruption events (TDEs) was reported in the literature. The light curve of these TDEs shows a re-brightening feature in the decline phase after the first peak, which then leads to a second flare. The re-brightening TDEs challenge the existing light curve fitting tools, designed to handle single flare. In this work, we present a model, aimed at reproducing the light curve of the re-brightening TDEs, based on the scenario that the consecutive flares are produced by the same star who experienced two partial disruptions (PTDEs). We also developed a fitting code from this model and applied it to two re-brightening TDEs: AT 2022dbl and AT 2023adr. The light curves of both TDEs are well-fitted. Finally, we forecast the time and peak brightness of the next flare for these two TDEs, so that the observers can prepare in advance and examine on our model.
\end{abstract}


\section{Introduction}
\label{SECT:Introduction}
A star that comes too close to a supermassive black hole (SMBH) shall be disrupted by the overwhelming tidal force from the SMBH, causing a tidal disruption event (TDE). The critical distance to the SMBH for a TDE is denoted as the tidal radius $r_{\rm t}$, an order-of-magnitude estimate of the tidal radius is given by $r_{\rm t} = r_*(M_{\rm BH}/m_*)^{1/3}$, where $M_{\rm BH}$, $m_*$, and $r_*$ are the SMBH mass, stellar mass, and radius, respectively. Such an event gives rise to a flare that can last months to years, making TDE a promising tool to detect the otherwise dormant SMBHs in the center of galaxies.

In the era before the Zwicky Transient Facility (ZTF; \cite{Bellm2019PASP..131a8002B}), the number of (candidate and confirmed) TDEs are accumulated at a pace of roughly 2 per year~\citep{Graham2019PASP,Gezari2021ARAA}: ever since the first TDE found in the archival data of the ROSAT all-sky survey~\citep{KB1999A&A}, X-ray and UV satellites have been the main avenues for hunting TDEs for more than one decade, then the ground-based optical surveys gradually catch up. The discovery rate is pushed up to a few dozens per year by the ZTF and will be further boosted with the upcoming survey facilities, such as the Legacy Survey of Space and Time (LSST) at the Vera Rubin Observatory (VRO) \citep{Ivezic2019ApJ...873..111I}. 

TDEs are featured by their characteristic light curves: the luminosity rises quickly to the peak, following either linear, quadratic or Gaussian rising. After the peak, the luminosity declines following an exponential or power law decay~\citep{vanVelzen+2021ApJ,HvVC2023}. To date, most of the observed TDEs exhibit only one flare, resulted from a single full- or partial disruption of the intruding star. This is consistent with the theoretical expectation. Ever since the work of \cite{Rees1988}, a commonly adopted assumption is that the disrupted star possesses essentially zero orbital energy (parabolic orbit), hence it can only have one close encounter with the SMBH.

However, as the observational data accumulates and the time baseline extends, some interesting TDEs have been observed: their light curves show re-brightening features in the declining phase of the first flare or long after the first flare. Currently, a handful of (candidate and confirmed) re-brightening TDEs have been reported in the literature: PS1-10adi~\citep{JWM2019}; ASASSN-14ko~\citep{Payne+2021ApJ}; AT 2018fyk~\citep{Wevers+2023ApJ}; AT 2020vdq~\citep{Somalwar2023arXiv}; AT 2022dbl~\citep{LinJiangWang+2024ApJ}; F01004-2237~\citep{Sun+2024arXiv}; AT 2019aalc~\citep{Veres+2024arXiv}; RX J133157.6–324319.7~\citep{HKG2022,MLR2023}; eRASSt J045650.3–203750~\citep{LMK2023,LRG2024}; Swift J023017.0+283603~\citep{Evans+2023NatAs,Guolo+2024NatAs}. We also note that a few TDEs presented by~\cite{Yao+2023ApJ} show re-brightening features: AT 2019baf, AT 2019ehz, AT 2020acka, AT 2021uqv. Although small in number, they already exhibit great diversity in the light curve properties. In the optical-selected TDE samples, the time interval between the first and second flares spans a range from a few hundred days (e.g., the sample of \cite{Yao+2023ApJ}) to a few years \citep{Somalwar2023arXiv} or even one decade~\citep{Sun+2024arXiv}. Another aspect is the brightness ratio between the two peaks: in some cases, the second flare is dimmer than the first flare, while in other cases, the second flare is brighter (e.g., AT 2020vdq, AT 2019aalc, F01004-2237). Lastly, the number of repeating flares from the same target also varies greatly, from a minimum of 2 to more than 20 (e.g., ASASSN-14ko).

The nature of this re-brightening or repeating phenomenon is still an open question, and there are many possible models. In the first model, this phenomenon is produced by repeated partial TDEs of the same star \citep{Somalwar2023arXiv,Wevers+2023ApJ,Liu+2024arXiv,Bandopadhyay+2024ApJ}. Contrary to full disruption, a partial TDE only strips away a fraction of mass from the star, and a remnant star will survive. The amount of stripped mass depends primarily on the penetration factor $\beta$ and stellar structure (\cite{GRR2013}, hereafter GRR2013). If the remnant star moves in an eccentric orbit, it can return to the pericenter periodically and produce repeating PTDEs. Then the time interval between the two peaks is mainly determined by the orbital period of the star. The repeating PTDEs might also be responsible for another interesting type of nuclear transient, the quasi-periodic eruptions (QPEs) observed in the X-ray band~\citep{WYM2022}, e.g. GSN 069~\citep{Miniutti+2019}, RXJ1301.9+2747~\citep{GMS2020}, eRO-QPE1, eRO-QPE2~\citep{Arcodia+2021}. QPEs have much shorter periods (hours to weeks) than the repeating flares observed in the UV/optical bands. Given the short time interval between the peaks, the star is tightly bound to the MBH, which is viable through the tidal break-up of hard binaries in the galactic center (also known as the Hills mechanism, \cite{Hills1988,Cufari+2022ApJ,Wevers+2023ApJ}). During the PTDE, the tidal perturbation excites stellar oscillation, and the tidal torques spin up the remnant star \citep{Ryu+2020ApJ,ChenDaiLiu2024arXiv,Bandopadhyay+2024ApJ}. Therefore, the response of the stellar structure to these perturbations determines the strength and the peak brightness of the subsequent disruption (detailed in Section~\ref{SECT:Discussion}).

In the second model, the two flares are full TDEs produced by two different stars following the tidal break-up of a binary \citep{Mandel_Levin2015ApJ}. In this model, only two flares can be produced because the stars are totally destroyed. Because the fallback rate is largely affected by the orbital parameters (as we will discuss later), the relative brightness and the time interval between the two flares are determined by the complicated three-body dynamics.

\cite{ChenDouShen2022ApJ} have proposed a third model which employs two-phase accretion. In this model, the first flare is powered by the fallback of tidal debris, and a temporally delayed accretion process is responsible for the second flare, which is dimmer than the first flare and generally does not follow the mass fallback rate due to the viscous delay operating in the accretion disk. This model has been successfully applied to AT 2019avd. \cite{CLG1990} have also used a viscous delayed $\alpha$-disk model~\citep{SS1973} to study the light curve of TDE.

Analyzing the observed TDE light curves could help us to reveal the physical mechanisms responsible for the phenomenon of re-brightening TDE. Currently, there are two published TDE light curve fitting tools in the literature: the Modular Open Source Fitter for Transients (\texttt{MOSFiT})~\citep{MOSFiT,MGR2019} and \texttt{TiDE}~\citep{TiDE}. They are aiming at extracting the physical parameters behind the TDEs, especially the mass of the SMBH, the mass of the disrupted star, and its orbital parameter. These tools adopt different physical models and assumptions: \texttt{MOSFiT} adopts a luminosity-dependent photosphere with black body spectral energy distribution (SED), while the emission model of \texttt{TiDE} adopts an accretion disk and a reprocessing layer. \texttt{MOSFiT} and \texttt{TiDE} have been used in many studies \citep{MGR2019,Gomez+2020,Nicholl+2020MNRAS,MR2021ApJ,HvVC2023,KV2023}. 

\texttt{MOSFiT} and \texttt{TiDE} share a common assumption that the intruding star is moving on parabolic orbit ($e=1$) before the disruption because the disrupted star is originated from the influence radius of the MBH\footnote{By definition, the stars residing within the influence radius are bound to the MBH, while those outside the influence radius are unbound.}~\citep{FR1976}. Therefore, the star can encounter the SMBH only once. One may try to fit the individual flares in the re-brightening TDEs but shall encounter the following problems: 1) the mass fallback rate of $e<1$ disruption, which is the case for re-brightening TDEs, is different from that of $e=1$ case adopted by \texttt{MOSFiT} and \texttt{TiDE} \citep{HSL2013MNRAS,PH2020ApJ,LMRR2023ApJ}. Hence, these tools can not model the individual flares accurately. 2) During the second flare, there might be some contribution from the first flare, which is currently not accounted for in \texttt{MOSFiT} and \texttt{TiDE}. 3) the time separation between the two flares are not modeled in \texttt{MOSFiT} and \texttt{TiDE}. Therefore, these tools are not suitable for fitting the re-brightening TDEs. 

In this work, we construct a new model to fit the light curve of re-brightening TDEs, assuming they are produced by the repeating PTDEs. Here, we focus on the optical light curve modeling, and leave the modeling of the X-ray light curve for future study. In Section~\ref{SECT:Model}, we first describe how to construct a composite mass fallback rate curve that consists of contributions from the two consecutive PTDEs. Then, this mass fallback rate curve is converted to the multi-band mock light curves adopting the luminosity-dependent photosphere model, which is also implemented in \texttt{MOSFiT}.
In Section~\ref{SECT:applications}, we apply our light curve model to two re-brightening TDEs: AT 2022dbl and AT 2023adr. Then we forecast the time and peak brightness of the next flare for these two TDEs. In Section~\ref{SECT:Discussion}, we discuss the caveats of our model and possible improvements that can be made in future development. Finally, we summarize this paper in Section~\ref{SECT:Summary}.

\section{Model}
\label{SECT:Model}

In this section, we describe the model for the re-brightening TDEs. We only consider the case that one single star experienced two partial TDEs since observations currently found most of the re-brightening TDEs only exhibit two flares.

\subsection{The mass fallback rate of $e=1$ disruption}
\label{SUBSEC:dmdE_parabolic}

We first explain how to get the mass fallback rate in the standard $e=1$ case, and this procedure mainly follows the one implemented in \texttt{MOSFiT}. Then, we derive the mass fallback rate for the $e<1$ case in the next subsection.

The construction of mass fallback rate is based on the hydrodynamic simulation of GRR2013. GRR2013 simulated the disruption of $1~M_{\odot}$ star by a $10^6~M_{\odot}$ SMBH, with various of penetration factor $\beta$ and a constant eccentricity $e=1$. The penetration factor $\beta$ is defined as the ratio of tidal radius $r_{\rm t}$ to the pericenter distance $r_{\rm p}$. The resultant debris energy distributions ($dm/d\epsilon$) for a set of discrete $\beta$'s, are obtained from the official website of \texttt{MOSFiT}\footnote{https://github.com/guillochon/MOSFiT} \citep{MOSFiT,MGR2019}. GRR2013 adopted two polytropic stellar models: $\gamma=4/3$ (suitable for $1~M_{\odot}\leq m_{*} \leq 15~M_{\odot}$) and $\gamma=5/3$ (suitable for $m_{*} \leq 0.3~M_{\odot}$ and $m_{*} \geq 22~M_{\odot}$). In the rest parts of the mass ranges, we follow the procedure implemented in \texttt{MOSFiT} to construct a hybrid fallback function (in this paper, we call the stars in these mass ranges hybrid stars). \cite{MGR2019} defined the fractional contributions, $g_{\rm frac}$, from the two $\gamma$ models as follows:
\begin{equation}
\label{eq:gfrac_1}
g_{\rm frac}=\frac{m_*/M_{\odot}-1}{0.3-1},~~~~~(0.3 < m_{*}/M_{\odot} <1)
\end{equation}

\begin{equation}
\label{eq:gfrac_2}
g_{\rm frac}=\frac{m_*/M_{\odot}-15}{22-15}.~~~~~(15 < m_{*}/M_{\odot} < 22)
\end{equation}
\noindent
Therefore, $g_{\rm frac} = 0$ is corresponding to $\gamma=4/3$ model, $g_{\rm frac} = 1$ is corresponding to $\gamma=5/3$ model. Then the mass fallback rate for hybrid star ($\dot{M}_{\rm hybrid}$) is obtained by linearly interpolating between the $\gamma=4/3$ and $\gamma=5/3$ fallback rate functions, $\dot{M}_{4/3}$ and $\dot{M}_{5/3}$ provided by GRR2013. $\dot{M}_{4/3}$ and $\dot{M}_{5/3}$ only depend on $\beta$, so the next step is to choose the appropriate $\beta$'s for these two functions. Note the range of $\beta$ for the partial disruption of the two $\gamma$ models are different: the partial disruption of $\gamma=4/3$ ($\gamma=5/3$) star occurs at $0.6<\beta_{4/3}<1.85$ ($0.5<\beta_{5/3}<0.9$). In practice, \texttt{MOSFiT} adopts a scaled penetration factor $b$ to describe the strength of disruption, and uses the following equations to translate $b$ to $\beta$ (and vice versa), for the partial disruption of polytropic stars,

\begin{equation}
\label{eq:b_to_beta_43}
\beta_{4/3} = 0.6 + 1.25b,~~~~~(0\leq b \leq 1,~\gamma=4/3)
\end{equation}

\begin{equation}
\label{eq:b_to_beta_53}
\beta_{5/3} = 0.5 + 0.4b.~~~~~(0\leq b \leq 1,~\gamma=5/3)
\end{equation}
\noindent
These equations indicate that $b=0$ corresponds to no disruption and $b=1$ corresponds to full disruption.

For the hybrid star, $\beta$ is obtained by linearly interpolating between $\beta_{4/3}$ and $\beta_{5/3}$, i.e. $\beta = \beta_{4/3} + (\beta_{5/3}-\beta_{4/3}) g_{\rm frac}$. After some simple algebra, we find 
\begin{equation}
\label{eq:map_b_hybrid}
b=\frac{\beta-0.6+0.1 g_{\rm frac}}{1.25-0.85g_{\rm frac}}.
\end{equation}
Note, by setting $g_{\rm frac}=0$ ($g_{\rm frac}=1$), equation~\ref{eq:map_b_hybrid} reduces to equation~\ref{eq:b_to_beta_43} (equation~\ref{eq:b_to_beta_53}). In the end, we can calculate the mass fallback rate of hybrid star with the following interpolation equation, 
\begin{equation}
\label{eq:Mdot_interpolation}
 \dot{M}_{\rm hybrid}(b) = \dot{M}_{4/3}(b) + [\dot{M}_{5/3}(b)-\dot{M}_{4/3}(b)] g_{\rm frac}   
\end{equation}

The mass fallback rate is scaled to the target black hole mass and star, using the scaling relation $\dot{M}_{\rm peak} \propto M_{\rm BH}^{-1/2} m_*^{2} r_*^{-3/2}$ given by GRR2013. Finally, we convert the mass fallback rate to the energy distribution of bound debris using the equation $dm/d\epsilon = \dot{M} / (d\epsilon/dt)$. The time derivative of $\epsilon$ can be derived from the Kepler's third law. The resulted energy distribution of the tidal debris is denoted as $(dm/d\epsilon)_{e=1}$. We could also compute the stripped mass during the partial TDE, 
\begin{equation}
\label{eq:Delta m}
\Delta m = 2\int_{-\infty}^{0} \left(\frac{dm}{d\epsilon}\right)_{e=1} d\epsilon,
\end{equation}
\noindent
because the $(dm/d\epsilon)_{e=1}$ function is symmetric to $\epsilon=0$.

\subsection{The mass fallback rate of $e<1$ disruption}
\label{SUBSEC:dmdE_eccentric}

The next step is to modify the mass fallback rate according to the orbital eccentricity of the disrupted star. Here we only consider the $e<1$ case, because the star capable of producing multiple PTDEs has to stay on an eccentric orbit. \cite{HSL2013MNRAS} have shown that the effect of reducing orbital eccentricity is shifting the $dm/d\epsilon$ curve toward the direction of negative energy with a displacement of $\epsilon_{\rm orb}$,

\begin{equation}
\label{eq:Shift_of_dmde}
\frac{dm}{d\epsilon}(\epsilon)= \left(\frac{dm}{d\epsilon}\right)_{e=1} (\epsilon-\epsilon_{\rm orb}).
\end{equation}
\noindent
The relation between orbital energy and eccentricity reads

\begin{equation}
\label{eq:E_orb(e)}
\epsilon_{\rm orb} = -\frac{\beta (1-e)}{2} q^{1/3} \Delta \epsilon,
\end{equation}
\noindent
where $q=M_{\rm BH}/m_*$ is the mass ratio between the SMBH and the star, and $\Delta \epsilon = GM_{\rm BH}r_*/r_{\rm t}^2$ is the typical energy scale of the $dm/d\epsilon$ distribution. Finally, we convert $dm/d\epsilon$ to mass fallback rate, i.e. $dm/dt = (dm/d\epsilon)(d\epsilon/dt)$, and denote it as $\dot{M}(t;M_{\rm BH},m_{*},b,\epsilon_{\rm orb})$.


The procedure of shifting $dm/d\epsilon$ (equation~\ref{eq:Shift_of_dmde}) is also justified by the results of \cite{LMRR2023ApJ}. They estimated the overall impact of tidal distortions on the star as it travels from the pericenter to the apocenter, which solely depends on the orbital eccentricity. They found that a $5\%$ deviations to the tidal deformation from cases in parabolic orbits occur when $e=e_{5\%}$. Hence, equation~\ref{eq:Shift_of_dmde} can be safely applied when $e>e_{5\%}$. According to Figure 1 of \cite{LMRR2023ApJ}, we find that $\log(1-e_{5\%})$ is roughly $-1.3$ (i.e. $e_{5\%}$ is roughly $0.95$), therefore the $e>e_{5\%}$ condition is always satisfied in the parameter space we are exploring (see the typical value of $e$ computed below).

In the light curve fitting procedure, we use the orbital period of the intruding star before the first close encounter, $P_{\rm orb}$, as a fitting parameter instead of $e$. Because $P_{\rm orb}$ is easier to read from the light curve, we can set a reasonable prior range for it. We note that after the first encounter, the orbital period of the remnant star will change due to the orbital energy gain or loss (detailed in Section~\ref{SUBSEC:Change_of_E_orb}) induced by the tidal encounter. Hence, the time interval between the two flare peaks should only be regarded as a rough estimate of $P_{\rm orb}$. It is also convenient to convert $P_{\rm orb}$ to $\epsilon_{\rm orb}$ according to Kepler's third law, 
\begin{equation}
\label{eq:Eorb(Porb)}
\epsilon_{\rm orb}= -\left( \frac{\pi G M_{\rm BH}}{\sqrt{2}P_{\rm orb}} \right)^{2/3}.
\end{equation}
\noindent
For completeness, we also derive the relation between $e$ and $P_{\rm orb}$. Inserting equation~\ref{eq:E_orb(e)} into equation~\ref{eq:Eorb(Porb)} results in

\begin{equation}
\label{eq:e(Porb)}
e = 1-\left( \frac{2\pi t_{\rm *,dyn}}{P_{\rm orb}} \right)^{2/3} \beta^{-1},
\end{equation}
\noindent
where $t_{\rm *,dyn}=\sqrt{r_*^3/(Gm_*)} \simeq 1.6\times 10^3~(r_*/R_{\odot})^{3/2} (m_*/M_{\odot})^{-1/2}~\rm{s}$ is the dynamic timescale of the star. Consider a typical case, where $t_{\rm dyn}\simeq 1.6\times 10^3 \rm{s}$, $P_{\rm orb}\simeq 100$ days, and $\beta\simeq 1$, we find $e\simeq 0.989$.

\subsection{Variation of the orbital energy of the remnant star}
\label{SUBSEC:Change_of_E_orb}

The remnant star that emerged from the partial disruption may lose or gain orbital energy, affecting the returning time and the mass fallback rate of the next disruption (c.f. equation~\ref{eq:Shift_of_dmde}). Therefore, taking this effect into account is important for accurate light curve modeling.

\cite{MGR2013} conducted the first study on this problem by performing Newtonian hydrodynamic simulations. They measured the orbital energy change after a partial disruption of a $\gamma = 4/3$ polytrope star, and found that the remnant star always gains orbital energy. The amount of energy gain monotonically increases with $\beta$, but never exceeds the surface escape velocity ($v_{\rm esc}$) of the original star. They also provided a fitting formula to calculate the orbital energy gain. \cite{GTG2015} performed General Relativistic simulations of PTDE, and obtained a similar conclusion that the remnant star gains orbital energy, though they used $\gamma=5/3$ polytrope star. Later, \cite{Ryu+2020ApJ} studied the same problem and also took the General Relativistic effect into account, but the star is modeled with realistic stellar structure. Contrary to the previous two studies, \cite{Ryu+2020ApJ} found that the orbital energy of the remnant star is not always increasing: the remnant star produced in the partial disruption of low-mass star actually loses orbital energy after a weak encounter, while for high-mass star, the remnant may also loses orbital energy even in some severe encounters. Recently, \cite{ChenDaiLiu2024arXiv} studied the orbital energy variation after a PTDE for both $\gamma = 4/3$ and $\gamma = 5/3$ stars, and found that there is a critical penetration factor, denoted as $\beta_{\rm t}$: the remnant star loses (gains) orbital energy if $\beta < \beta_{\rm t}$ ($\beta > \beta_{\rm t}$).

\begin{figure}
    \centering
    \includegraphics[width=0.8\linewidth]{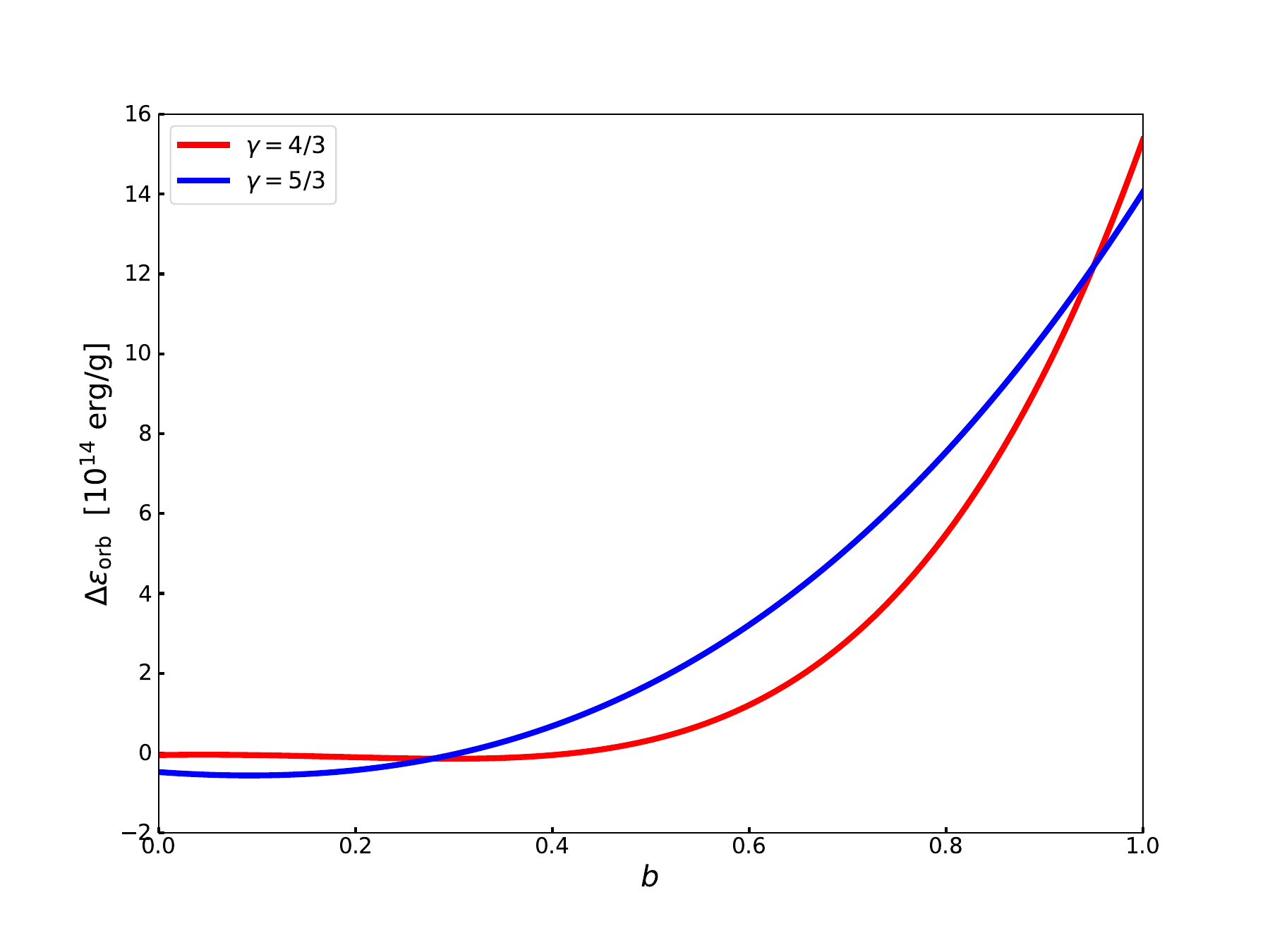}
    \caption{The dependence of $\Delta\epsilon_{\rm orb}$ on the scaled penetration factor $b$ for the two $\gamma$ stellar models (indicated by color), based on equation 9 of \cite{ChenDaiLiu2024arXiv}. We have transformed $\beta$ in their equation 9 to $b$, using equation~\ref{eq:map_b_hybrid}.
    }
    \label{fig:DeltaE_vs_b}
\end{figure}

\cite{ChenDaiLiu2024arXiv} provided two fitting formulae (their equation 9) to compute the orbital energy changes $\Delta \epsilon_{\rm orb}$ of the remnant stars (both $\gamma = 4/3$ and $\gamma = 5/3$ stellar models) after every PTDE, which only depends on $\beta$. In order to obtain $\Delta \epsilon_{\rm orb}$ for the hybrid stars, we firstly convert the variable $\beta$ in the original fitting formulae to $b$ (see the plot in Figure~\ref{fig:DeltaE_vs_b}), then compute the orbital energy change using
\begin{equation}
\label{eq:Delta E_orb interpolation}
\Delta\epsilon_{\rm orb}(b) = \Delta\epsilon_{\rm orb, 4/3}(b) + [\Delta\epsilon_{\rm orb, 5/3}(b) - \Delta\epsilon_{\rm orb, 4/3}(b)]\times g_{\rm frac}.  
\end{equation}
\noindent
We also impose an upper limit on $\Delta\epsilon_{\rm orb}$: it should not exceed $v_{\rm esc}^2/2$. As a result, the orbital energy of the remnant star is $\epsilon_{\rm orb,new} = \epsilon_{\rm orb}+\Delta\epsilon_{\rm orb}$. Accordingly, the new orbital period is calculated using Kepler's third law.

\subsection{The composite mass fallback rate}
\label{SUBSEC:composite mass fallback rate}

In this section we construct the composite mass fallback rate, which consists of two subsequent PTDEs produced by the same star. We start with a brief description of the procedure and key parameters.

A star with initial mass $m_{*,0}$ and radius $r_{*,0}$ approaches the SMBH on an eccentric orbit with initial scaled penetration factor $b_0$ ($0<b_0<1$, the corresponding $\beta_0$ is derived from equation~\ref{eq:map_b_hybrid}) and the rest-frame orbital period $P_{\rm orb,0}$ (the corresponding orbital energy $\epsilon_{\rm orb,0}$ is derived from equation~\ref{eq:Eorb(Porb)}). It passes the pericenter at the time of $t=t_{\rm disrupt}$, and the first partial TDE ensues. After that, a remnant star with mass $m_{*,1}$ and radius $r_{*,1}$ continues its journey around the SMBH. The stellar mass $m_{*,1} = m_{*,0} - \Delta m$, where $\Delta m$ is computed from equation~\ref{eq:Delta m}, and $r_{*,1}$ is computed from the mass-radius relation of zero-age-main-sequence (ZAMS) star \citep{Tout+1996MNRAS}. Accordingly, the new tidal radius for disrupting the remnant star is computed by $r_{\rm t,1} = r_{*,1}(M_{\rm BH}/m_{*,1})^{1/3}$, and the new penetration factor is $\beta_1 = r_{\rm t,1} / r_{\rm p,1}$ and the associated $b_1$ is obtained with equation~\ref{eq:map_b_hybrid}. Hydrodynamic simulations have found that the pericenter of this new orbit is almost the same as the initial orbit \citep{Ryu+2020ApJ,ChenDaiLiu2024arXiv}. Therefore we take $r_{\rm p,1} = r_{\rm p,0}$ for simplicity. The orbital energy of this remnant star is also updated by an amount of $\Delta\epsilon_{\rm orb}$ (equation~\ref{eq:Delta E_orb interpolation}), i.e., $\epsilon_{\rm orb,1} = \epsilon_{\rm orb,0} + \Delta \epsilon_{\rm orb}$. Then, we use Kepler's third law to compute the orbital period for the remnant star, $P_{\rm orb,1}$.

With these parameters, we could obtain the mass fallback rate of the individual PTDEs, denoted as $\dot{M}_{0}(\Tilde{t};M_{\rm BH},m_{*,0},b_0,\epsilon_{\rm orb,0})$ and $\dot{M}_{1}(\Tilde{t};M_{\rm BH},m_{*,1},b_1,\epsilon_{\rm orb,1})$, respectively. In both functions, $\Tilde{t}$ is the time variable and the other quantities are parameters. Note $\Tilde{t}$ measures the elapsed time since the corresponding disruption, i.e., $\Tilde{t}=0$ when the (remnant) star passes the pericenter, while $t$ measures the date of the observations. Therefore, $\Tilde{t}=t-t_{\rm disrupt}$ in $\dot{M}_0$ and $\Tilde{t}=t-(t_{\rm disrupt}+P_{\rm orb,1})$ in $\dot{M}_1$, because the remnant star passes the pericenter at the time $t_{\rm disrupt}+P_{\rm orb,1}$.

We assume the first light of the first PTDE emerges at the time when stream-stream collision occurs, denoted as $t_0=t_{\rm disrupt} + 1.5 t_{\rm min,0}$, where $t_{\rm min,0}$ is the orbital period of the most bound debris in the first PTDE, both of them are given in unit of days. Since the remnant star takes $P_{\rm orb,1}$ to return to the pericenter and give rise to the second PTDE, the light curve of the second PTDE starts at $t_1=t_{\rm disrupt} + P_{\rm orb,1} + 1.5 t_{\rm min,1}$, where $t_{\rm min,1}$ is the orbital period of the most bound debris in the second PTDE. The final composite mass fallback rate $\dot{M}_{\rm comp}$ is the summation of $\dot{M}_0$ and $\dot{M}_1$.

\begin{figure}
    \centering
    \includegraphics[width=0.4\linewidth]{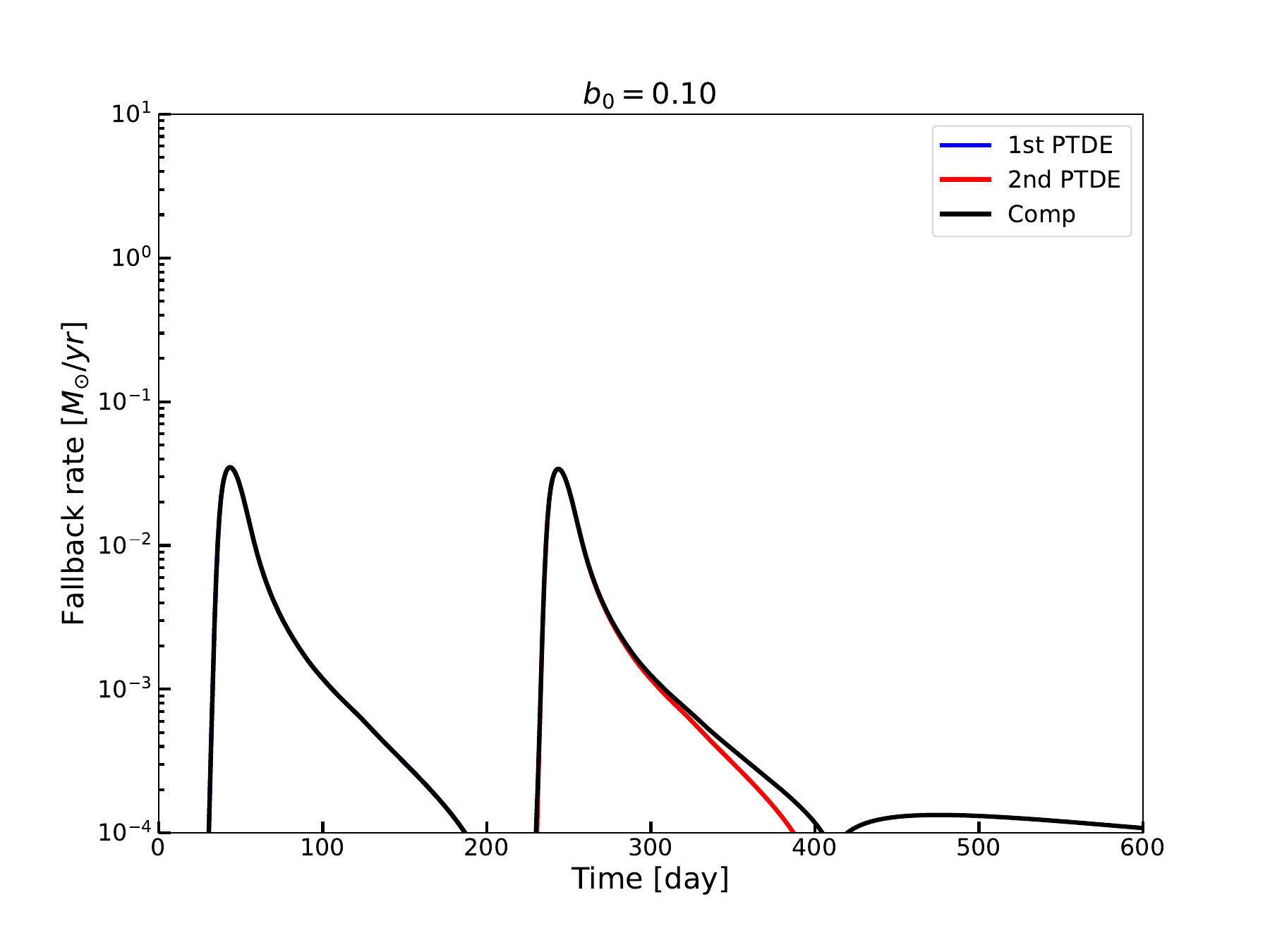}
    \includegraphics[width=0.4\linewidth]{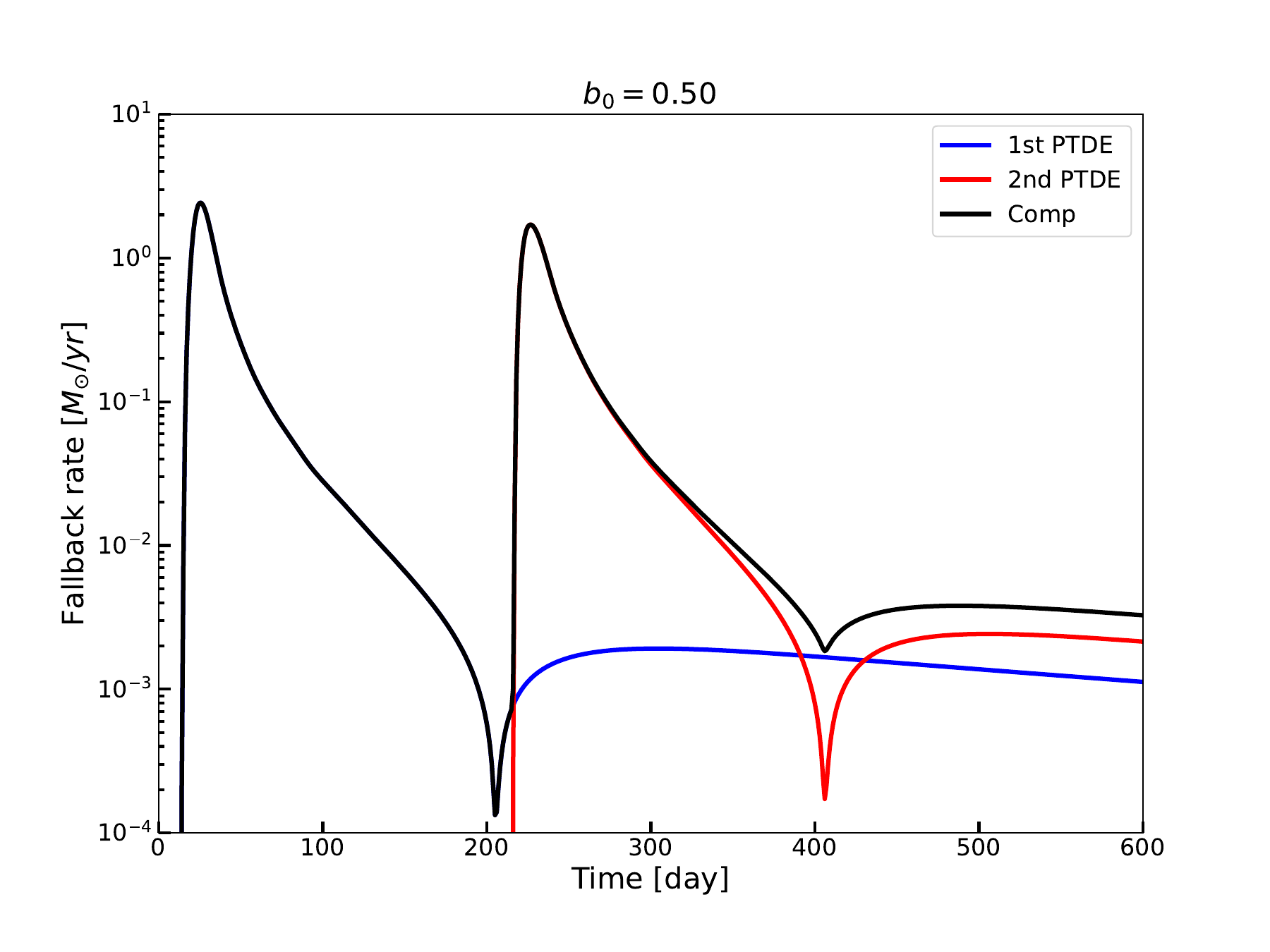}
    \includegraphics[width=0.4\linewidth]{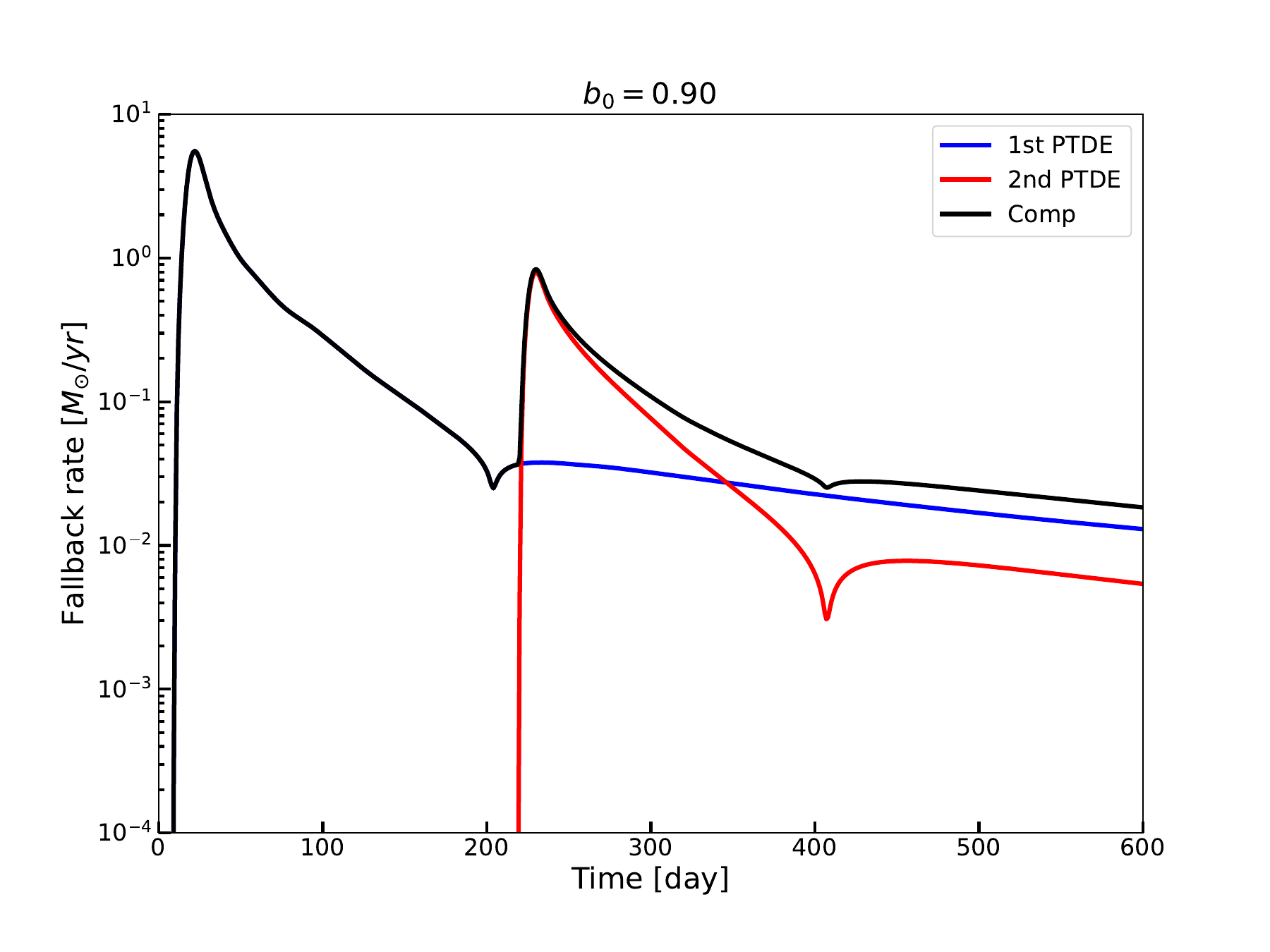}
    \includegraphics[width=0.4\linewidth]{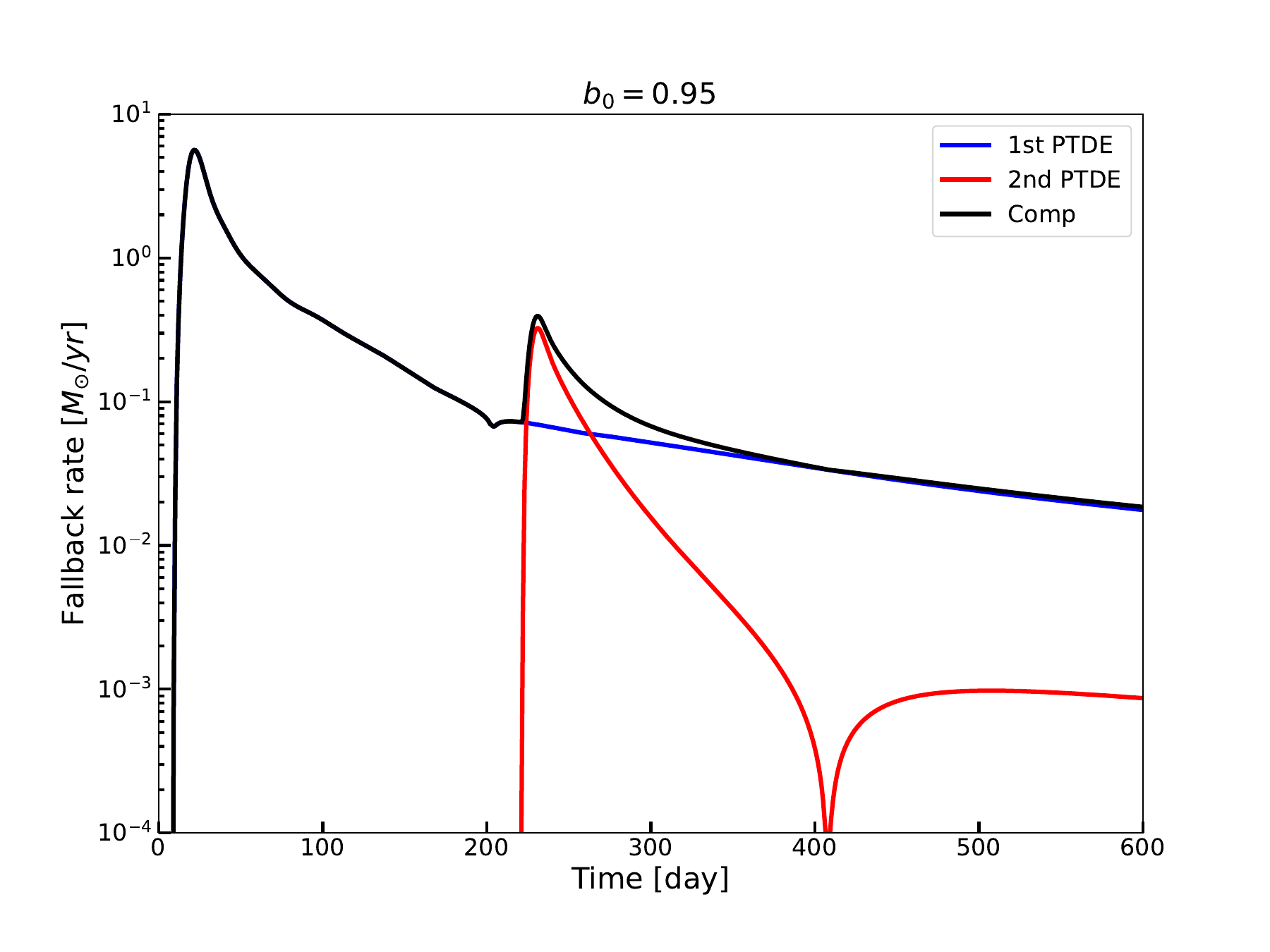}
    \caption{The composite mass fallback rate with various of $b_0$. The blue and red curves represent the mass fallback rate of the first ($\dot{M}_0$) and second ($\dot{M}_1$) PTDE, respectively. The starting time of $\dot{M}_0$ is shifted to $t_0$, and the starting time of $\dot{M}_1$ is shifted to $t_1$ (see the main text). The black curve represents the composite mass fallback rate of the two consecutive PTDEs. In all these plots, we set $M_{\rm BH}=10^6~M_{\odot}$, $m_{*,0}=1~M_{\odot}$, $P_{\rm orb,0}=200$ days, $t_{\rm visc}=0$ day and $t_{\rm disrupt}=0$.
    }
    \label{fig:FallbackRate_b}
\end{figure}

Figure~\ref{fig:FallbackRate_b} gives four examples of the composite mass fallback rate made by two consecutive PTDEs to demonstrate the effect of the initial scaled penetration factor $b_0$ on the mass fallback rate. In this figure $b_0$ increases from the top left panel to the bottom right panel, and so do the peak mass fallback rates of the two PTDEs. It also shows that the ratio of two peak mass fallback rates increases with $b_0$: the two peaks are nearly the same when $b_0=0.1$, while in the case of $b_0=0.95$, the first peak is almost one order of magnitude higher than the second peak. Another feature we want to emphasize is the decay phase after the second peak. During this period, the materials stripped in the first PTDE are still falling back (the blue curve), effectively enhancing the composite fallback rate (the black curve) during the second peak. As a result, the fallback rate after the second peak decays slower than it would be in an solitary PTDE (the red curve). This effect is more significant in the high $b_0$ cases.

The accretion rate $\dot{M}_{\rm acc}(t)$ that actually powers the luminosity should defer from the fallback rate, due to the viscous delay of the accretion. Following the procedure in \texttt{MOSFiT}, we introduce a viscous timescale $t_{\nu}$ to control the viscous delay, and calculate $\dot{M}_{\rm acc,i}(\Tilde{t})$ with

\begin{equation}
\label{eq:M_acc}
\dot{M}_{\rm acc,i}(\Tilde{t}) = \frac{1}{t_{\nu}}
\left( e^{-\Tilde{t}/t_{\nu}} \int_0^{\Tilde{t}} e^{\Tilde{t'}/t_{\nu}} \dot{M}_{\rm i}(\Tilde{t'}) d\Tilde{t'} \right)
\end{equation}
\noindent
\citep{MGR2019}. Increasing $t_{\nu}$ could prolong the rising and declining timescales of the light curve, and suppress the peak luminosity. Note, only in the limit of $t_{\nu} \rightarrow 0$ shall the accretion rate equal the mass fallback rate.

\subsection{Multi-band Light Curve Modeling}
\label{SUBSEC:LightCurveModel}

In this section we describe how to generate the multi-band light curve from the mass fallback rates. In practice, we first compute the bolometric light curve for the individual PTDEs ($L_{\rm bol,0}$ and $L_{\rm bol,1}$) from the associated viscously delayed mass fallback rate ($\dot{M}_{\rm acc,0}$ and $\dot{M}_{\rm acc,1}$). Then compute the magnitude of the two flares ($m_{\rm band,0}$ and $m_{\rm band,1}$) in every observed bands, using the radiation model described below. Finally, the composite light curve is obtained by

\begin{equation}
\label{eq:compositeMag}
m_{\rm band} = -2.5\log(10^{-0.4 m_{\rm band,0}} + 10^{-0.4 m_{\rm band,1}}).
\end{equation}

We adopts the luminosity-dependent photosphere model implemented in \texttt{MOSFiT} to convert the mass fallback rate to multi-band light curve. We give a brief introduction to this radiation model, more details can be found in \cite{MGR2019}.

The bolometric luminosity is computed by $L_{\rm bol,i}(t) = \eta \dot{M}_{\rm acc,i}(t) c^2$, where $i$ marks the individual PTDEs (0 stands for the first PTDE, 1 stands for the second PTDE), $\eta$ is the radiation efficiency, $c$ is the speed of light and $\dot{M}_{\rm acc,i}(t)$ is obtained by equation~\ref{eq:M_acc}. The range of $\eta$ is set between $10^{-4}$ and $0.1$, the upper limit is the typical radiation efficiency of an AGN disk. The lower limit comes from minimum value achievable for the eccentric accretion disk model \citep{ZLK2021}.

The optical spectral energy distribution (SED) of the individual flares is generally described by black body radiation with effective temperature $T_{\rm eff,i}$ \citep{Gezari2021ARAA}. The total bolometric luminosity is the product of the area of the emitting surface, $4\pi R_{\rm ph,i}^2$, and the energy flux per unit area $\sigma_{\rm SB} T_{\rm eff,i}^4$, where $R_{\rm ph,i}$ is the so called photosphere radius, $\sigma_{\rm SB}$ is the Stefan-Boltzmann constant. After some simple algebra, the expression for the effective temperature reads

\begin{equation}
T_{\rm eff,i} = \left ( \frac{L_{\rm bol,i}}
   {4\pi \sigma_{\rm SB} R_{\rm ph,i}^2} \right )^\frac{1}{4}.
\label{Eq:T_eff}
\end{equation}
\noindent
Then the SED is $F_{\rm i}(\lambda)=B_{T_{\rm eff,i}}(\lambda) R_{\rm ph,i}^2 / D_{\rm L}^2$, where $B_{T_{\rm eff,1}}(\lambda)$ is the Planck function and $D_{\rm L}$ is the luminosity distance.

Many observations of TDEs have found that the effective temperature varies little near the peak and tend to increase at late times. In order to model this SED behavior in \texttt{MOSFiT}, a power-law scaling relation between the $R_{\rm ph,i}$ and $L_{\rm bol,i}$ is adopted,

\begin{equation}
R_{\rm ph,i} = R_{\rm ph0} a_{\rm p,i}(L_{\rm bol,i}/L_{\rm Edd})^l,
\label{Eq:R_ph}
\end{equation}
\noindent
where $R_{\rm ph0}$ is a normalization of the photosphere radius, $a_{\rm p,i}$ is the semimajor axis of the material corresponding to the maximum fallback rate, and $L_{\rm Edd}$ is the Eddington luminosity. Substitute this relation into equation~\ref{Eq:T_eff} results in $T_{\rm eff,i}\propto L_{\rm bol,i}^{(1-2l)/4}$, hence in the special case of $l=1/2$, the effective temperature will not vary with time.

Finally, a conversion from bolometric luminosity to the AB magnitude in different bands is performed. This is done with the $F(\lambda)$ and the filter transmission function $T(\lambda)$, using the following equation

\begin{equation}
m_{\rm AB} = -2.5\log \int \lambda F_{\rm i}(\lambda) T(\lambda) d\lambda
+2.5\log \int \frac{T(\lambda)}{\lambda} d\lambda -2.408.
\label{Eq:AB_mag}
\end{equation}

\section{Apply our model to the re-brightening TDEs}
\label{SECT:applications}

This section tests our model with two TDEs having re-brightening features, namely AT 2022dbl and AT 2023adr. \cite{LinJiangWang+2024ApJ} confirmed that AT 2022dbl was produced by repeating PTDEs, and the authors claimed that the third flare is expected to recur in the 2026 (see also our forecasting in Section~\ref{SUBSECT:forecast}). AT 2023adr was recently reported as a repeating PTDE \citep{LQ2024TNSAN}, and the expected third flare should occur earlier than AT 2022dbl. Thus, these two cases provide good opportunities for further validating our model, if the third flare do occur at the predicted time and brightness.

We also note there are other newly discovered re-brightening TDEs: AT 2020vdq \citep{Somalwar2023arXiv}, and AT 2019aalc \citep{Veres+2024arXiv}. In these cases, the second flare is even brighter than the first flare. Note, AT 2019aalc took place in AGN, which might introduce more complexities in the light curve modeling, due to the underlying AGN variability. Currently, our model can not handle these features (see also Section~\ref{SUBSECT:FlareBrightness}), hence we exclude these TDEs from our analysis. For the TDEs presented in~\cite{Yao+2023ApJ} that have re-brightening features (AT 2019baf, AT 2019ehz, AT 2020acka, AT 2021uqv), we are already late for observing their third flares, so we also exclude them from our analysis.

The multi-band light curve data of AT 2022dbl was retrieved from \cite{LinJiangWang+2024ApJ}. The \textit{Swift}/UVOT $U$- and $B$-band photometry is not very reliable for this target (Zheyu Lin, private communication), we exclude the data points in these two bands in the fitting procedure. \cite{LinJiangWang+2024ApJ} has derived the host MBH mass: $\log(M_{\rm BH}/M_{\odot}) = 6.40 \pm 0.33$ with the $M_{\rm BH}$-$\sigma$ relation of \cite{KormendyHo2013ARAA}.

The light curve data of AT 2023adr is retrieved from the ALeRCE ZTF Explorer\footnote{https://alerce.online/}, which performs the $g$- and $r$-band differential photometry on the target TDE. Then we correct the Milky Way extinction for this target, using $R_V=3.1$ and $E(B-V)=0.011$ from the SFD extinction map \citep{SF2011ApJ}. The resultant extinction is $0.042$ mag in $g$-band and $0.029$ mag in $r$-band.

We use the python package \texttt{emcee} \citep{emcee2013PASP} to optimize the fitting, by maximizing the logarithmic likelihood 
\begin{equation}
\log p = -\frac{1}{2} \sum_{i} \left[ \frac{x_i^2}{\sigma_i^2} + \log (2\pi \sigma_i^2) \right],
\end{equation}
\noindent
where $x_i$ is the difference between the $i$th observation and model prediction, and $\sigma_i$ takes the error of the $i$th observation. Maximizing the above logarithmic likelihood is equivalent to minimizing the $\chi^2 = \sum x_i^2/\sigma_i^2$ value.

The prior distribution of some common parameters are given in Table~\ref{table-paras}. For AT 2022dbl, the prior distribution of $P_{\rm orb,0}$ is between 650 and 690 days, $t_{\rm disrupt}$ is between MJD 59550 and 59600. For AT 2023adr, the prior distribution of $P_{\rm orb,0}$ is between 340 and 370 days, $t_{\rm disrupt}$ is between MJD 59850 and 59910. 
We employ 100 walkers in the MCMC fitting and run for 10,000 iterations. The results of our interested fitting parameters and the corresponding reduced chi-square value ($\chi_{\rm red}^2$) are summarized in Table~\ref{table-target-info}. 

\begin{table}[htbp]
\caption{The fitting parameters used in the light curve modeling.
\label{table-paras}}
\begin{center}
\begin{tabular}{cccc}
  \hline
  Parameter & Prior Distribution Type & Min  &  Max   \\
  \hline
  $M_{\rm BH}/M_{\odot}$  & Log   & $10^6$  & $10^8$ \\
  $m_{*}/M_{\odot}$       & Kroupa & $0.1$ & $5$ \\
  $b$ (scaled penetration factor)  & Flat & $0.5$ & $0.95$ \\
  $\eta$                  & Log   & $10^{-4}$ & $0.1$ \\
  $R_{\rm ph0}$           & Log    & $10^{-2}$ & $10^2$ \\
  $l$                     & Flat   & $0$ & $2$ \\
  $t_{\nu}$/days          & Log    & $10^{-1}$ & $10$ \\
  \hline
\end{tabular}
\end{center}
\textbf{NOTES:}~The first column gives the name of the parameters. The second column indicate the type of prior distribution for each parameters: ``Flat" means the prior is uniformly sampled from the value range; ``Log" means the prior is logrithmically uniformly sampled in the value range; ``Kroupa" means the stellar mass are sampled from the Kroupa initial mass function. The third and fourth columns give the allowed range for each parameter.
\end{table}

\begin{table}[htbp]
\caption{Basic information of the re-brightening TDEs and the corresponding fitting results}
\label{table-target-info}
\begin{center}
\begin{tabular}{c|ccc|cccc|c}
  \hline
  IAU name & R.A. & Decl. & redshift & $\log(M_{\rm BH}/ M_{\odot})$ & $m_{*,0}/M_{\odot}$ & $b_0$ & $P_{\rm orb,0}$/day & $\chi_{\rm red}^2$ \\
  \hline
  AT 2022dbl  & $12^{h}20^{m}45.097^{s}$ & $49^{\circ}33^{'}4.86^{''}$ & $0.028$  & $6.91^{+0.02}_{-0.01}$ & $1.00^{+0.00}_{-0.00}$ & $0.63^{+0.00}_{-0.00}$ & $680.44^{+0.16}_{-0.17}$ & 14.0\\
  AT 2023adr  & $14^{h}36^{m}19.847^{s}$ & $32^{\circ}23^{'}16.44^{''}$ & $0.131$  & $7.18^{+0.04}_{-0.06}$ & $0.90^{+0.03}_{-0.03}$ & $0.80^{+0.02}_{-0.02}$ & $346.97^{+1.20}_{-1.21}$ & 1.3 \\
  \hline
\end{tabular}
\end{center}
\textbf{NOTES:} $m_{*,0}$, $b_0$ and $P_{\rm orb,0}$ are the stellar mass, scaled penetration factor and rest-frame orbital period before the first PTDE, respectively. These three quantities shall vary after the first PTDE, see details in Section~\ref{SUBSEC:composite mass fallback rate}. We use the 16th and 84th percentiles of the posterior distribution to indicate the $1\sigma$ uncertainty in all fitted parameters. The last column is the reduced chi-square value, calculated with the data points involved in the fitting procedure.
\end{table}

\begin{figure}
    \centering
    \includegraphics[width=0.45\linewidth]{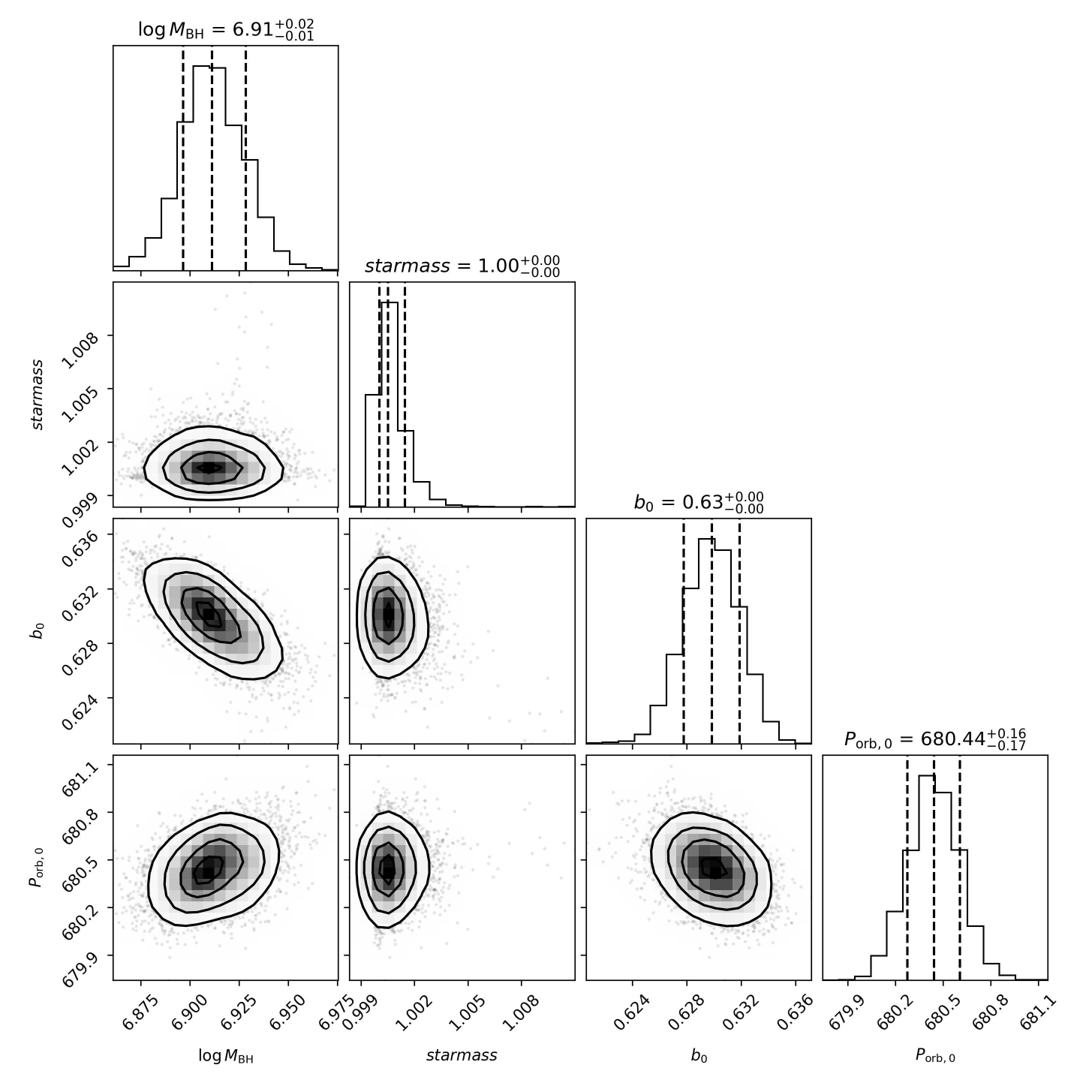}
    \includegraphics[width=0.45\linewidth]{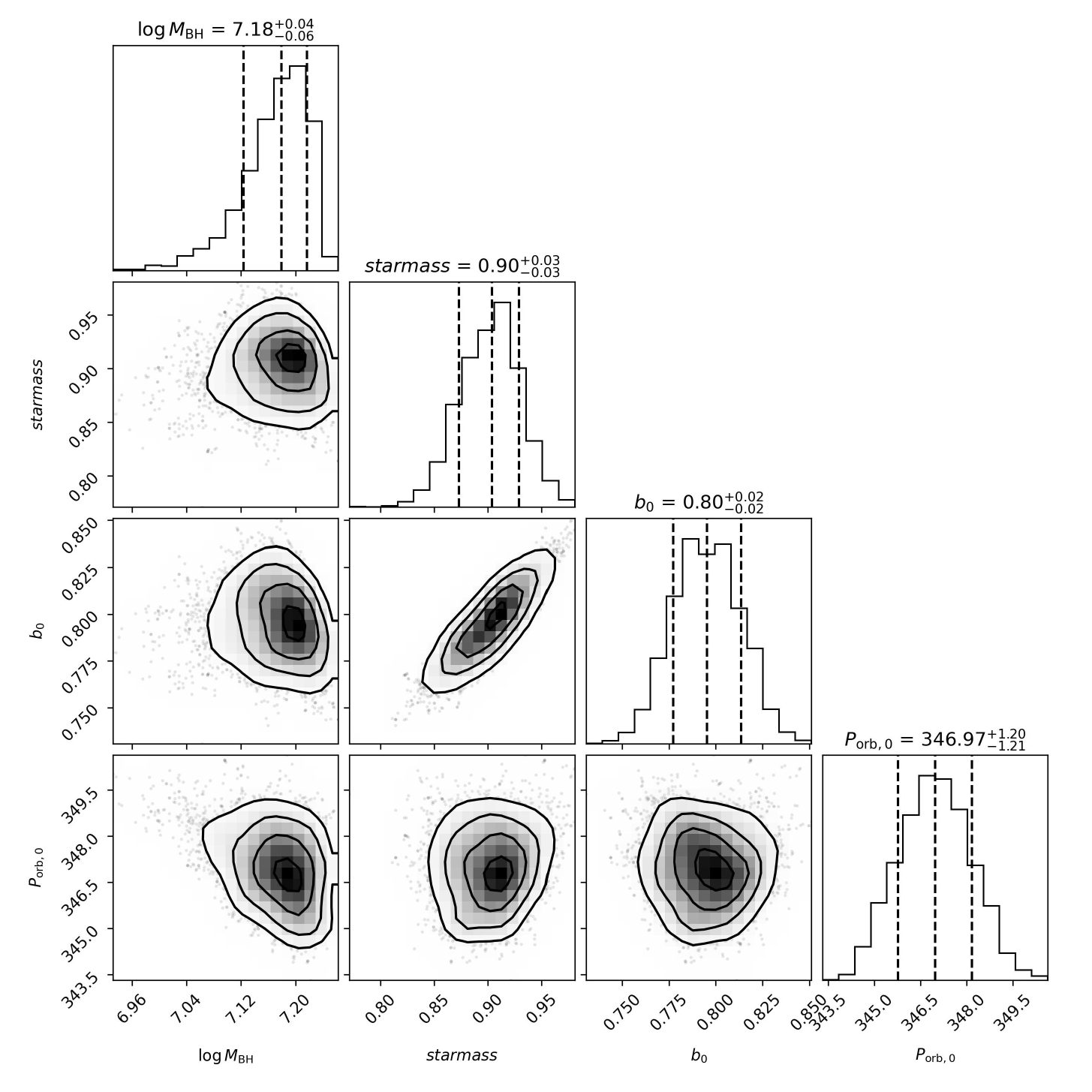}
    \caption{Posterior distributions of $\log (M_{\rm BH}/M_{\odot})$, $m_{*,0}$, $b_0$ and $P_{\rm orb,0}$, for AT 2022dbl (left) and AT 2023adr (right).
    }
    \label{fig:Result_posterior}
\end{figure}

Figure~\ref{fig:Result_posterior} presents the posterior distributions of $M_{\rm BH}$, $m_{*,0}$, $b_0$ and $P_{\rm orb,0}$ which interest us the most. The $M_{\rm BH}$ of AT 2022dbl obtained from the light curve fitting ($\log(M_{\rm BH}/M_{\odot})\simeq 6.91$) is higher than that derived by \cite{LinJiangWang+2024ApJ}, but the difference is within 1 dex. We note that in the other TDEs that have $M_{\rm BH}$ measurement, the values obtained from the TDEs are generally not exactly matching with the values derived from BH mass scaling relations, the agreement can only be achieved within $\sim 1$ dex (e.g. see the comparison in \cite{MGR2019} and \cite{KV2023}, but also see \cite{ZLK2021} who gets a relatively good agreement). There is no $M_{\rm BH}$ measurement for the host galaxy of AT 2023adr.

\begin{figure}
    \centering
    \includegraphics[width=0.48\linewidth]{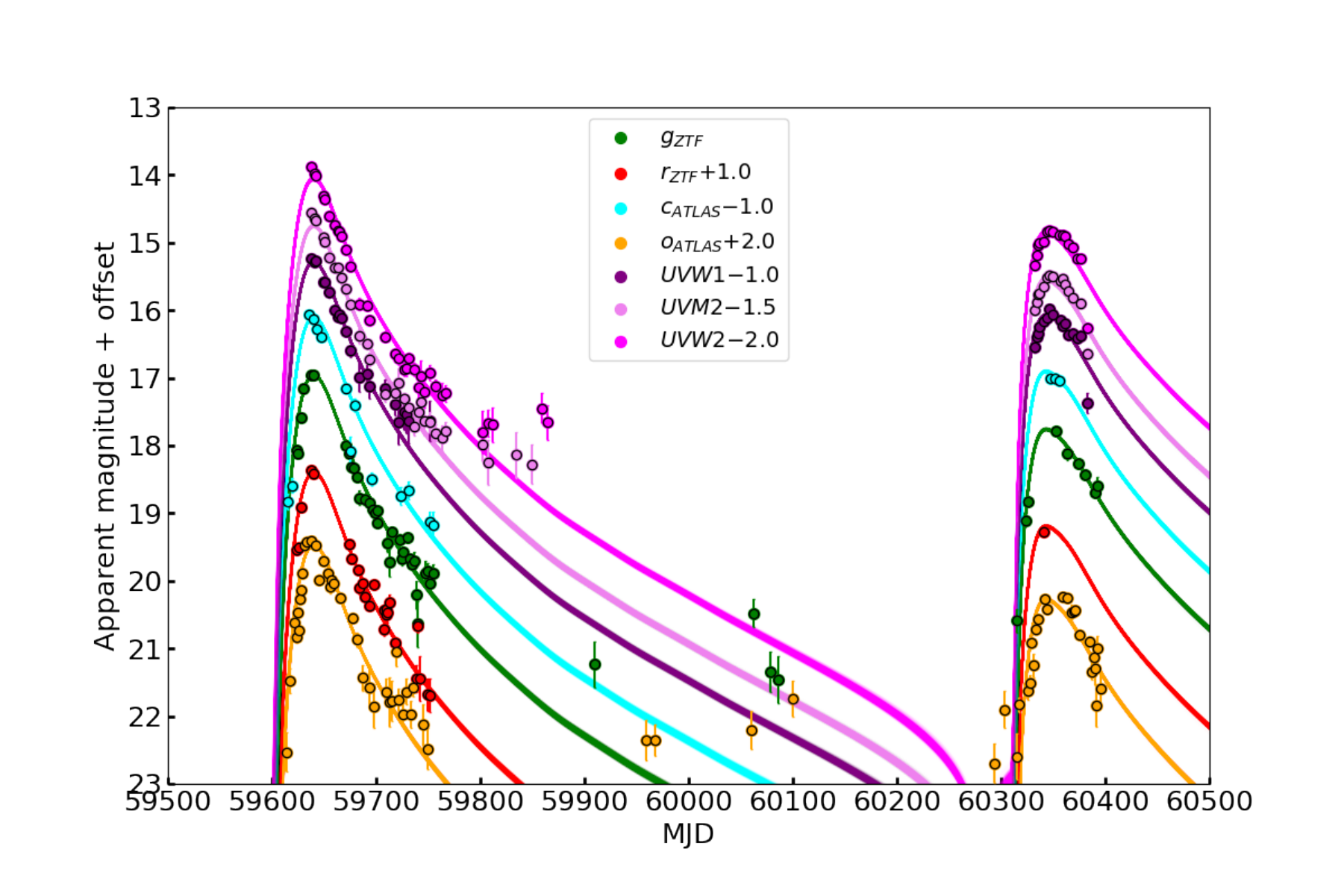}
    \includegraphics[width=0.48\linewidth]{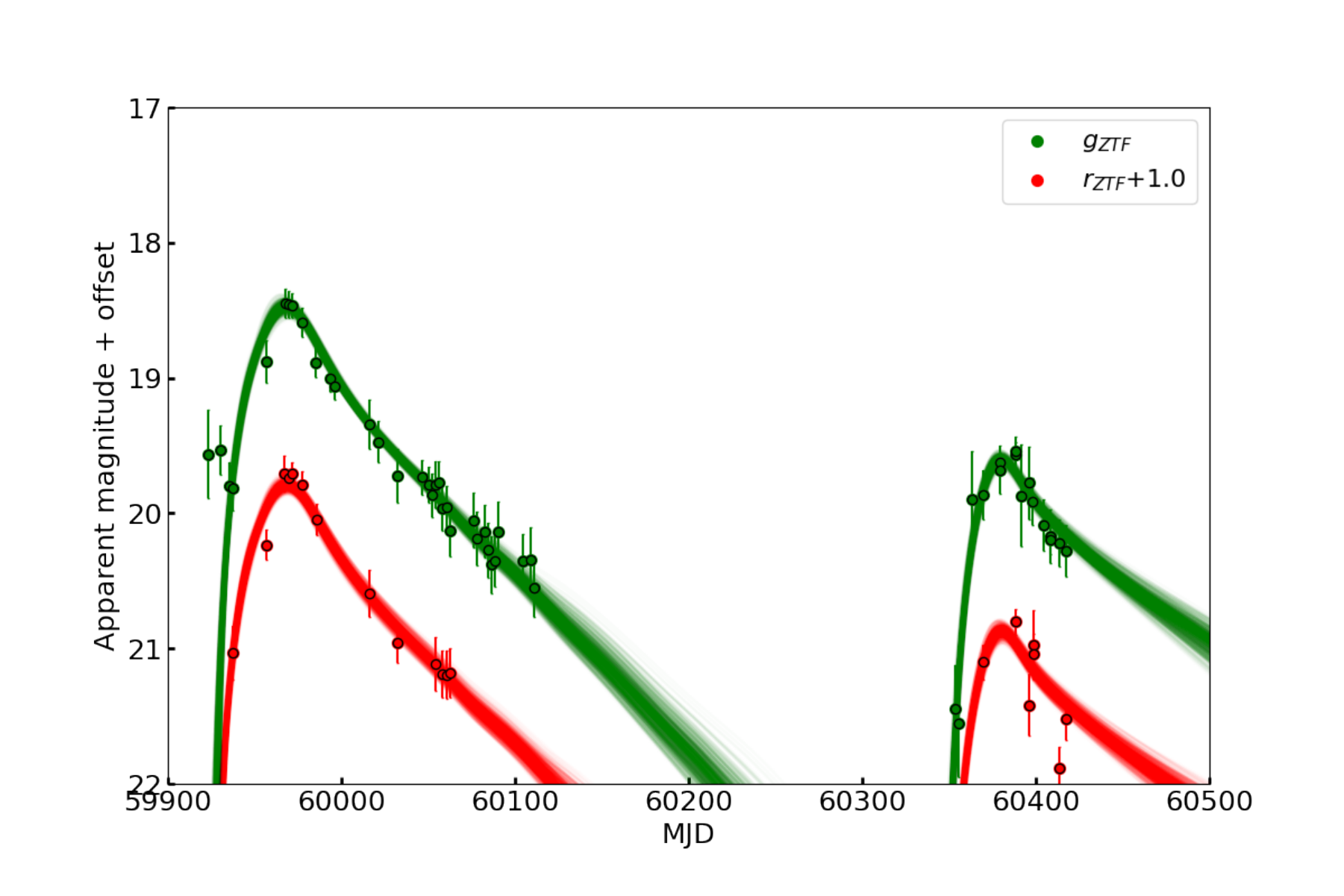}
    \caption{Fitting results for AT 2022dbl (left) and AT 2023adr (right). The dots represent the observations, and the curves represent the mock light curves generated by the fitting code.
    }
    \label{fig:Result_Light_Curve}
\end{figure}

Figure~\ref{fig:Result_Light_Curve} shows the observed light curve and the mock light curve generated by the fitting code, for visual comparison. The mock light curves are generated with parameters randomly sampled from the last 1,000 iterations. The agreement between the mock and observed light curves are generally good. However, we also notice some deviations of the observed data from the mock light curves:

\begin{itemize}
    \item AT 2022dbl: the deviation of the observed data in the $UVW2$ and $UVM2$ bands from the mock light curves starts about 100 days after the first peak. The ZTF $g$ and ATLAS $o$ bands also exhibit signatures of flattening in the decline phase of the first flare, but the beginning time is not obvious, due to the sparse sampling after MJD 59800. This feature resembles the late stage flattening observed in the UV bands in the other TDEs \citep{vVSM2019ApJ}, but occurs much earlier. This feature is likely attributed to the formation of accretion disk \citep{MB2020MNRAS,MvVN2024MNRAS}. Since the contributions from different mechanisms for the UV/optical radiations should evolve with time (e.g. stream-stream collision in the early stage and disk accretion in the late stage), this flattening feature might be handled with an evolving $\eta$, or with more physically motivated radiation models (also see Section~\ref{SUBSECT:RadiationModel}). We have excluded the data points between MJD 59800 and 60300 from the light curve fitting procedure. Including those data points would result in a black hole mass $\sim 0.2$ dex higher than the value reported in Table~\ref{table-target-info}, but the influences on the other fitting parameters are small.
    \item AT 2023adr: the earliest two $g$-band observation points are much brighter than the mock light curve.  Recently, early bumps in the rising light curves have been found in several TDEs and may be a commmon feature~\citep{CPLA2023A&A,HuangJiangZhu2024ApJ,WWJ2024,FAMH2024ApJ}. The earliest two $g$-band observation points in the light curve of AT 2023adr might be the endpoint of such early bump. Unfortunately, there is no data earlier than these two points (we also searched the archival data of ASASSN and ATLAS) to confirm or rule out this conjecture. Since our model can not handle the early bump, we have excluded these two points in the fitting procedure, otherwise, the mock light curve of the first flare becomes systematically brighter than the observations.
\end{itemize}

\subsection{Forecast the next flare}
\label{SUBSECT:forecast}
To identify the nature of re-brightening TDEs, a third flare from the same target is essential in distinguishing the repeated PTDEs from the double TDEs \citep{Mandel_Levin2015ApJ} and the two-phase evolution scenarios \citep{ChenDouShen2022ApJ}.

\begin{figure}
    \centering
    \includegraphics[width=0.48\linewidth]{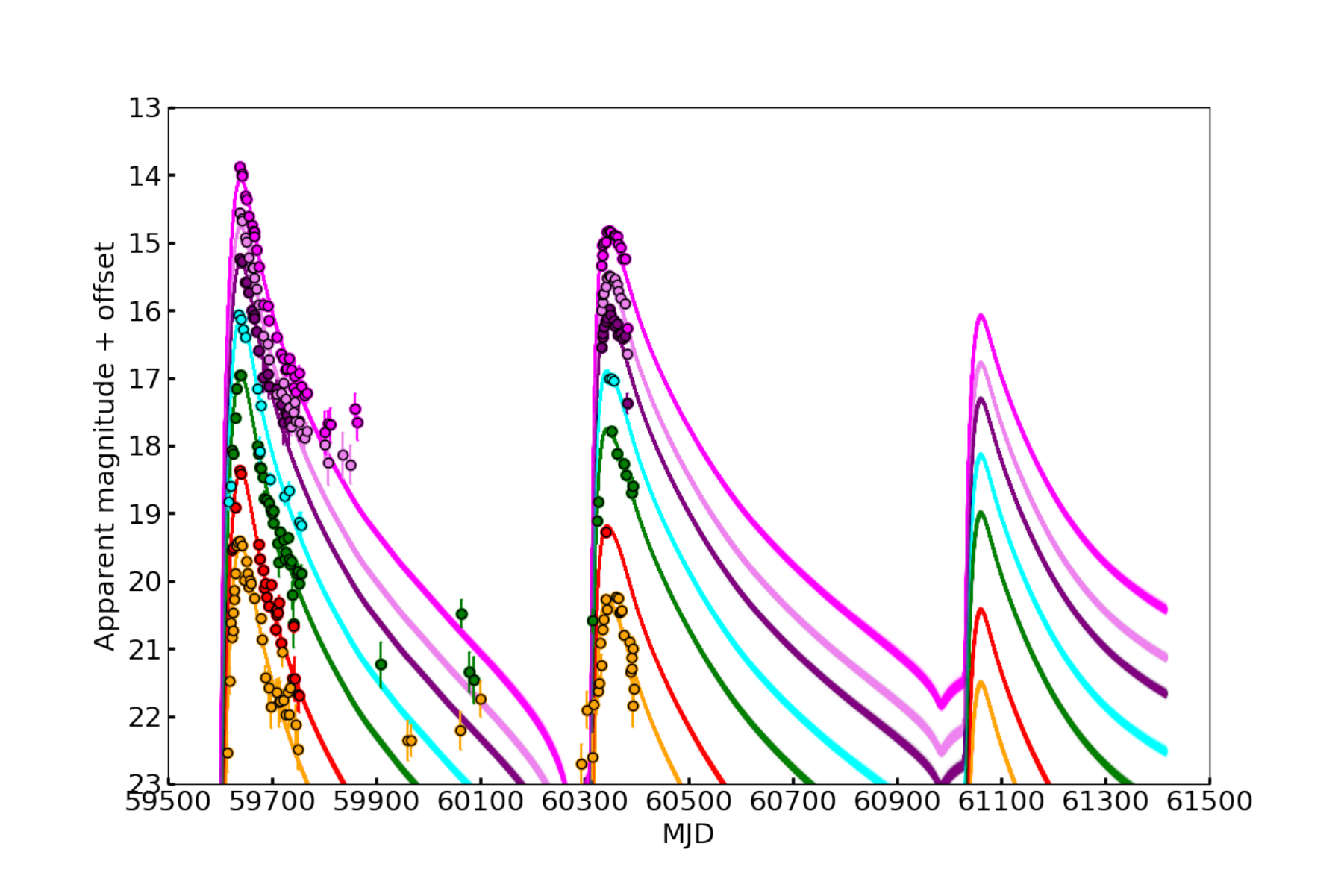}
    \includegraphics[width=0.48\linewidth]{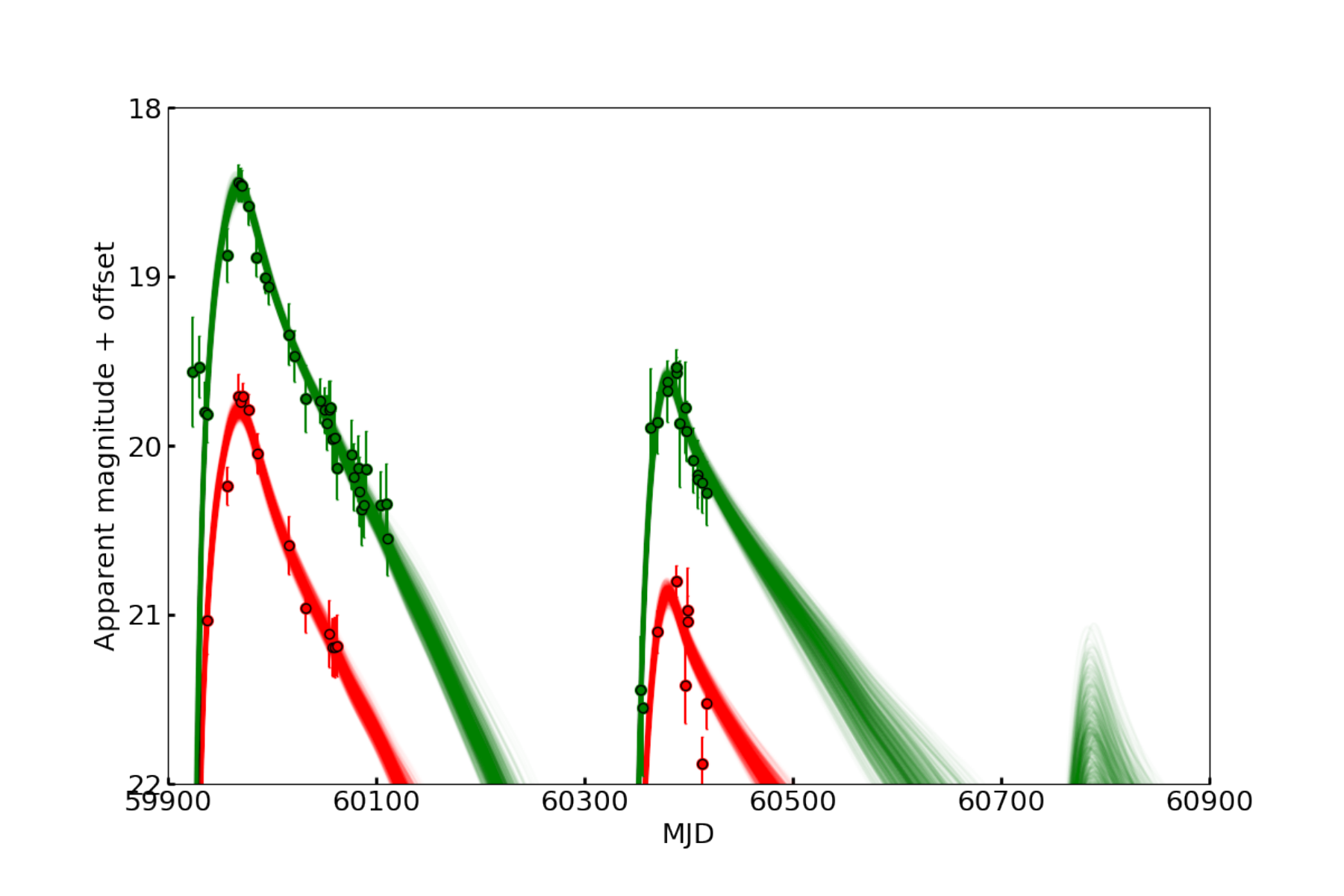}
    \caption{Forecast the next flare for AT 2022dbl (left) and AT 2023adr (right). The color and offset settings of different bands are the same as Figure~\ref{fig:Result_Light_Curve}.
    }
    \label{fig:Result_forecast}
\end{figure}

Despite of modeling the light curve of re-brightening TDEs that has been observed, we could also forecast the time and multi-band light curves of the next flare for the two targets (Figure~\ref{fig:Result_forecast}), based on the fitting results obtained from the first and second flares. The corresponding peak time and brightness in individual bands are given in Table~\ref{table-forecast}.

The third flare of AT 2023adr is expected to occur in April of 2025. The large $b_{0}$ causes a very low remnant star mass after the first two PTDEs, and in the third PTDE the stripped mass is $\Delta m_* = 0.007^{+0.006}_{-0.004}~M_{\odot}$. Together with the large distance modulus of AT 2023adr, the peak brightness of the third flare is very faint ($\sim 22$ mag). Such a faint flare may be detectable by deep time-domain surveys such as 2.5 meter Wide Field Survey Telescope~\citep{Wang+2023SCPMA}, or specifically monitored by medium or large telescopes.

In the case of AT 2022dbl, the moderate $b_0$ ensures that there is enough remnant mass to power the third flare ($\Delta m_* =0.083^{+0.004}_{-0.002}~M_{\odot}$). The third flare is expected to occur in January of 2026, and it is sufficiently bright to be observed by both ground and space telescopes, providing a better case to test our model prediction.

These predictions are valuable for TDE studies. The early phase data could provide constraint on the radiation mechanisms of TDEs, but is currently lacking of multi-band observations, and especially the spectrum in rising phase. Using our predicted time information, observers could prepare for the observation of next flare in advance. According to the predicted brightness, observers could decide which instrument should be used to do spectroscopy and/or photometric observations, raising the probability of obtaining high quality data.

\begin{table}[htbp]
\caption{Forecast the peak time and brightness of the third flare}
\label{table-forecast}
\begin{center}
\begin{tabular}{c|c|ccc|cc|cc}
  \hline
  IAU name & MJD & $UVW2$ & $UVM2$ & $UVW1$ & ZTF-$g$ & ZTF-$r$ & ATLAS-$c$ & ATLAS-$o$  \\
  \hline
  AT 2022dbl  & $61061.3^{+0.5}_{-0.5}$ & $18.1^{+0.0}_{-0.0}$ & $18.3^{+0.0}_{-0.0}$ & $18.3^{+0.0}_{-0.0}$ & $19.0^{+0.0}_{-0.0}$ & $19.4^{+0.0}_{-0.0}$ & $19.1^{+0.0}_{-0.0}$ & $19.5^{+0.0}_{-0.0}$\\
  AT 2023adr  & $60787.7^{+4}_{-3}$ & $22.1^{+0.5}_{-0.6}$ & $22.1^{+0.4}_{-0.5}$ & $22.0^{+0.3}_{-0.4}$ & $21.9^{+0.3}_{-0.3}$ & $22.1^{+0.3}_{-0.3}$ & $21.9^{+0.3}_{-0.3}$ & $22.2^{+0.3}_{-0.3}$ \\
  \hline
\end{tabular}
\end{center}
\textbf{NOTES:} The time and peak brightness of the third flare, based on the parameters fitted from the previous two flares. We use the 16th and 84th percentiles of the posterior distribution to indicate the $1\sigma$ uncertainty in all quantities.
\end{table}

\section{Discussion}
\label{SECT:Discussion}

In the previous section, our light curve model has successfully reproduces light curve of two TDEs which have re-brightening features. And we have forecast the properties of the third flare. However, it is possible that the third flare will not exactly follow the forecasting, because our model is simplified and adopts many assumptions. In this section, we discuss the caveats of our model and possible improvements that can be done in the future.

\subsection{Relative brightness of the consecutive flares}
\label{SUBSECT:FlareBrightness}

In our model, the second flare shall always be dimmer than the first flare, because we adopt the mass-radius relation of ZAMS stars, therefore the stellar radius decreases as a consequence of mass loss after every PTDE (i.e. $r_*\propto m_*^{\alpha}$, with $\alpha=0.8$ for $m_*<1M_{\odot}$ and $\alpha=0.6$ for $m_*>1M_{\odot}$), and so do the tidal radius ($r_{\rm t}\propto r_* m_*^{-1/3} \propto m_*^{\alpha-1/3}$) and $\beta$ for subsequent disruption. Consequently, the peak mass fallback rate and the amount of stripped mass decrease according to equations A1 and A3 of GRR2013, respectively.

Our model is clearly not appropriate for AT 2020vdq, F01004-2237 and AT 2019aalc, in which the second flare is brighter. In our framework, a brighter second flare is feasible only if $\beta$ increases in the next PTDE. According to the definition of $\beta$, there are two possible ways to achieve this goal: 1) reducing $r_{\rm p}$, equivalent to reducing the orbital angular momentum; 2) increasing $r_{\rm t}$, resulted from the expansion of the remnant star. We discuss these two possibilities in the following part.

Numerical simulations have demonstrated that the orbital angular momentum of the remnant star is invariant \citep{Ryu+2020ApJ,ChenDaiLiu2024arXiv}, other mechanisms that can change the orbital angular momentum of the remnant star are needed. The two-body relaxation process operating in the host star cluster of the MBH can gradually modify the orbital angular momentum of every member star, potentially decreasing the $r_{\rm p}$ of the remnant star. However, it is also possible that $r_{\rm p}$ increases when the remnant star return to the pericenter. Besides, for a star deeply buried in the gravitational potential well of the SMBH, the two-body relaxation process becomes extremely inefficient in altering the orbital angular momentum within one orbital period, because the relaxation timescale is many orders of magnitude longer than the orbital period. Hence, it is not likely that the brighter second flare is driven by the dynamical increase of $\beta$.

The process of PTDE resembles the fast mass transfer on dynamic timescales between binary stars, which could be treated as an adiabatic process \citep{HW1987,DBE2013,GHW2010ApJ,Ge+2015ApJ,GWC2020ApJ,GTC2024}. In this process, the stellar radius shrinks if the removed layer is radiative; on the contrary, an expansion in the stellar radius will ensue for convective layers. This behavior is also observed by ~\citet{Liu+2024arXiv}, who utilized the stellar evolution code \texttt{MESA} \citep{MESA} to examine the response of stars to adiabatic mass stripping (the mass is taken away by enhanced stellar wind). However, the picture becomes completely different, when tidal energy injection is considered. In the numerical simulations of \cite{Liu+2024arXiv}, a solar type star experienced repeatedly partial disruption by the MBH, with various of initial $\beta$ values. The tidal forces from the black hole induce oscillation in the remnant star \citep{MGR2013,ChenDaiLiu2024arXiv,Liu+2024arXiv,SPH2024MNRAS}. The energy stored in the oscillation modes shall eventually dissipate into thermal energy inside the star \citep{KumarGoodman1996ApJ}. If the injected thermal energy can not be radiated away within one orbital period, the remnant star will experience expansion in radius and a decrease in the mean density~\citep{Podsiadlowski1996MNRAS}, making the star more vulnerable to tidal disruption. If the stellar radius of the remnant star expands after the first PTDE, then the next PTDE can exhibit a brighter flare: $\beta$ increases with the stellar radius; hence the second disruption is more violent than the first one. \cite{Liu+2024arXiv} claimed that their findings might be responsible for the light curve of AT 2020vdq.

\cite{Bandopadhyay+2024ApJ} simulated a termination-age-main-sequence (TAMS) star being partially disrupted on grazing orbit around the MBH repeatedly, and found the stripped mass $\Delta m_*$ decreases in the successive PTDEs, which is attributed to the influence of the compact stellar core. However, the peak mass fallback rates ($\sim \Delta m_* / t_{\rm peak}$) are almost constant, because the effect of decreasing $\Delta m_*$ is roughly canceled out by the shortening of $t_{\rm peak}$, which is caused by the spinning up of the remnant star (the other effect of spin is discussed in Section~\ref{SUBSECT:TimeSeparation}). Unfortunately, they did not study the repeated PTDE of TAMS star with higher $\beta$. \cite{Bandopadhyay+2024ApJ} also studied the case of ZAMS star, the situation is a little bit complicated: on grazing orbit, the $\Delta m_*$ increases in the successive PTDEs; on the contrary, if the star has experienced a significant mass loss in the first PTDE, then the mass fallback rate of the second one would be lower.

Finally, we note that the model of \cite{Liu+2024arXiv} has difficulty reproducing the light curve of progressively dimming PTDEs, like the two targets presented in this work. While in the work of \cite{Bandopadhyay+2024ApJ}, only ZAMS stars could produce both types of repeated PTDEs (either brighter or dimmer second flare, depending on initial $\beta$).

\subsection{Time separation between the consecutive flares}
\label{SUBSECT:TimeSeparation}

The time interval between the consecutive flares is primarily determined by the initial orbital period. So far, we have only taken into account the orbital energy variation induced by the PTDE process (Section~\ref{SUBSEC:Change_of_E_orb}). However, there are other possible mechanisms that could also affect the time of the subsequent flares. In the following, we make a brief discussion on these mechanisms.

First, two-body relaxation process, despite of altering the orbital angular momentum, could also alter the orbital energy and period of the remnant star orbiting around the MBH. In the previous section, we mentioned that the orbital angular momentum variation caused by the two-body relaxation process is negligible based on the timescale argument. The energy relaxation timescale is even longer than the angular momentum relaxation timescale \citep{BW1976,CK1978}. Therefore, we can safely ignore the orbital energy variation caused by the two-body relaxation process from our modeling.

Second, the mass fallback rates measured by GRR2013 are all based on non-spinning stars. However, the spin of the (remnant) star can also alter the mass fallback rate. \cite{Golightly+2019ApJ} studied the full disruption of a spinning star, and found the tidal radius depends on the stellar spin: comparing to a non-spinning star, a spinning star on a prograde orbit (spin angular momentum is parallel to the orbital angular momentum) has a larger tidal radius. In contrast, a retrograde orbit (spin angular momentum is anti-parallel to the orbital angular momentum) reduces the tidal radius. \cite{Bandopadhyay+2024ApJ} extended the work of \cite{Golightly+2019ApJ} to repeated PTDEs, starting from a non-spinning star. The tidal torque from the central MBH induced spin angular momentum to the remnant star \citep{Ryu+2020ApJ,SPH2024MNRAS}. Consequently, \cite{Bandopadhyay+2024ApJ} found that the spin of the remnant star increases with the number of PTDEs, which causes the $t_{\rm peak}$ of the mass fallback to become shorter in the subsequent PTDEs. This effect could reduce the time interval between the two flares, if all the other parameters are kept the same. In this case, our model actually underestimates $P_{\rm orb,0}$, because a larger $P_{\rm orb,0}$ is needed to reproduce the observed time interval.

Another issue is that the star involved in the first PTDE might also possess high spin. The star capable of producing short period repeated PTDEs is generated by the Hills mechanism. After the tidal break-up of the binary star, one star becomes tightly bound to the MBH, while the other one is ejected. The orbital period of the bound star is determined by the initial orbital period of the binary system \citep{Pfahl2005ApJ,Cufari+2022ApJ}. To place a star on an orbital period of a few hundred days via Hills mechanism, the semi-major axis of the binary system must be very small. Consider an extreme case where a binary system consists of two $1 M_{\odot}$ stars with semi-major axis $a_{\rm b}=0.005$ AU, is tidally broken up by a $10^6 M_{\odot}$ MBH, the orbital period $P_{\rm orb,0}$ of the bound star is about $80$ days (using equation 4 of \cite{Pfahl2005ApJ}). Note in this extreme case, $a_{\rm b} \simeq 1 R_{\odot}$, the two stars are almost tidally locked (the spin period of the member star equals to the orbital period of the binary). Then the angular velocity of the bound star ($\Omega = 3.98\times 10^{-4}$ rad/s) is a significant fraction of its break-up spin ($\Omega_{\rm br} = \sqrt{GM_{\odot}/R_{\odot}^3} \simeq 6.27\times 10^{-4}$ rad/s). For less extreme cases (longer $P_{\rm orb,0}$), the $\Omega$ value is lower but may still change the mass fallback rate significantly. Unfortunately, a quantitatively assessment of the impact on the fitting results is currently unfeasible. There is no systematic study that simulates a grid of models with different stellar spin parameters (amplitude and orientation) and provides the mass fallback rate templates. Therefore, we cannot incorporate the effect of stellar spin into our model. We defer this issue to future work.

\subsection{Radiation Model}
\label{SUBSECT:RadiationModel}

In our light curve modeling, we use the same $\eta$, $R_{\rm ph,0}$, $l$ and $t_{\rm visc}$ for both flares (c.f. equation~\ref{eq:M_acc} and \ref{Eq:R_ph}), for the purpose of reducing the number of fitting parameters (currently 10 parameters) so that the fitting procedure could quickly converge. But there is no physical reason for this treatment.

From the observational perspective, in the case of AT 2022dbl, \cite{LinJiangWang+2024ApJ} found that the spectrum of the two flares are highly similar in the broad H$\alpha$ emission line, $\sim 4400-5200$\AA~and $\sim 4100$\AA~features. This indicates not only the two flares are produced by the same star (with the same chemical composition), but also the underlying radiation mechanism that generate these emission lines should be the same. Hence we think our treatment is acceptable, at least for AT 2022dbl.

The next issue is related to the radiation model itself. There are numerous physical mechanisms that could contribute to the luminosity of a TDE, such as nozzle shock and  spiral shocks near the MBH (pericenter shock), stream-stream collision (apocenter shock), stream-disk collision, and the accretion of a gas disk~\citep{Shiokawa+2015ApJ,Piran+2015ApJ,Bonnerot+Lu2022MNRAS,Ryu+2023ApJ.957.12R,Steinberg+Stone2024Natur,Huang+2024ApJ}. The luminosity should be dominate by the various shocks at early stages, then stream-disk collision joins after the formation of gas disk, and finally accretion takes over when the mass fallback rate significantly decreases. Therefore, a comprehensive description of luminosity should sum up all energy sources, with their relative contributions evolving over time. However, there is currently no consensus on which mechanisms should be dominant. For instance, \cite{Ryu+2023ApJ.957.12R} examined the role of pericenter and apocenter shocks in circularizing the debris and powering the luminosity. Although these shocks are not efficient in circularizing the debris, they provide sufficient energy to power the observed peak luminosity. Conversely, \cite{Steinberg+Stone2024Natur} found that the nozzle shock and stream-disk collision could efficiently circularize the fallback debris and are responsible for the pre-peak light curve. They also show that the accretion luminosity is subdominant to the shocks, but it is of the same order of magnitude. The discrepancy may be due to the initial conditions and the physical assumptions adopted by different authors.

Our primary goal is not to study these complex radiation mechanisms. Therefore, we adopt the radiation model implemented in \texttt{MOSFiT}. Although this is a simple phenomenological model, it works well in many previous works. In \texttt{MOSFiT}, the luminosity is converted from $\dot{M}_{\rm acc}$ via a constant radiation efficiency $\eta$. \cite{MGR2019} has mentioned that, as a fitting parameter, $\eta$ enables the conversion formula to accommodate various radiation mechanisms, ranging from shocks ($\eta$ less than one percent) to the accretion at the innermost stable circular orbit (ISCO) of a maximally spinning black hole ($\eta=0.4$). In this simplified conversion formula, $\eta$ should already accounts for all the processes that contribute to the luminosity.

Numerical simulations found that the luminosity before and around the peak is likely powered by the stream shocks~\citep{Ryu+2023ApJ.957.12R,Huang+2024ApJ}  and/or stream-disk collision~\citep{Steinberg+Stone2024Natur}. Our fitting results indicate this should be the case for the first flare, because the values of the fitted radiation efficiency ($\eta_{\rm fit}$, see Table~\ref{table-params2}) are much smaller than $0.1$ (the typical value in AGN accretion disk), but roughly align with the values for shock dissipation \citep{Jiang_2016ApJ,ZLK2021}. We also note that the viscous timescale is shorter than the fallback timescale $t_{\rm min,0}$ (see Table~\ref{table-params2}). This means $\dot{M}_{\rm acc}$ closely follows the $\dot{M}_{\rm fb}$, the growth of a gas disk might be too slow to affect the main body of the first flare, i.e. the role of accretion and stream-disk collision should be minor.

The situation for the second flare is more intricate: a gas disk could form during the first PTDE and it might not be exhausted by the time of the second PTDE. As a result, the luminosity of the second flare should also account for contribution from accretion process. Apart from accretion, if the gas disk is dense enough, the fallback stream of the second PTDE could disintegrate inside the disk: there will be no stream-stream collision, and stream-disk collision may become the dominant energy source. On the contrary, if the gas disk becomes diluted due to accretion and expansion, then the radiation mechanism of the second flare should resemble the first flare. In the following part, We examine the impacts of accretion and stream-disk collision on our results one by one.

In the case of AT 2022dbl, if the luminosity of the second flare is indeed powered by both shocks ($L_{\rm s}$) and accretion ($L_{\rm a}$), their relative intensity can be estimated as follows. We assume the flattening part after the first flare establishes the baseline of the accretion brightness ($m_{\rm a}$) and it does not evolve with time. We find in $g$ band, $m_{\rm a}$ is approximately 2.1 mag fainter than the peak brightness of the second flare ($m_{\rm peak}$). From the equation $m_{\rm peak}-m_{\rm a} = -2.5\log[(L_{\rm a}+L_{\rm s})/L_{\rm a}]$, we obtain the relative intensity $L_{\rm a}/L_{\rm s}$ as,

\begin{equation}
\frac{L_{\rm a}}{L_{\rm s}}=\frac{1}{10^{-0.4(m_{\rm peak}-m_{\rm a})}-1}.
\label{eq:Ld_vs_LS}
\end{equation}
\noindent
We find $L_{\rm a}/L_{\rm s} \simeq 0.169$. This value should be regarded as an upper limit, because there are marginal evidences that $L_{\rm a}$ declines with time (see Figure~\ref{fig:Result_Light_Curve}).


Regarding AT 2023adr, the flattening of the light curve was not captured in observations. If we consider the first data point of the second flare ($g=21.48$ mag) as the baseline of accretion plateau (because it is fainter than the last data point of the first flare), then the peak of the second flare ($g=19.61$ mag) is 1.87 mag brighter than the hypothetical plateau. Following the same analysis, we find $L_{\rm a}/L_{\rm s} \simeq 0.218$. This ratio is also an upper limit, because the hypothetical plateau might be fainter than the assumed value, and it may also decline with time.

We also estimated the radiation efficiency of stream-disk collision ($\eta_{\rm SD}$) in Appendix~\ref{Appendix-eta_SD} and listed the results in Table~\ref{table-params2}. By comparing the radiation efficiencies, we find that in AT 2022dbl, the luminosity from stream-disk collision between the stream of the second PTDE and the gas disk should be subdominant: $\eta_{\rm SD}$ is only one third of $\eta_{\rm fit}$. While in AT 2023adr, $\eta_{\rm SD}$ is much larger than $\eta_{\rm fit}$. If the stream-disk collision is indeed the dominant energy source in both flares of AT 2023adr, in order to reproduce the observed light curve, the fallback rate of both PTDEs should be reduced and our current model may overestimate the stellar mass. If the stream-disk collision luminosity only dominates in the second flare, then only the mass fallback rate of the second PTDE should be reduced, and we may underestimate the scaled penetration factor $b_0$ (the influence of $b_0$ on the relative strength of the two peaks is shown in Figure~\ref{fig:FallbackRate_b}). As a result, the third flare should be fainter than our prediction. Lastly, the latter case also implies that adopting a common $\eta$ for both flares may be inappropriate for AT 2023adr.

The late-time flattening of the AT 2022dbl light curve is not captured by the currently adopted model (Figure~\ref{fig:Result_Light_Curve}), but could be reproduced by accretion disk model \citep{MvVN2024MNRAS}. Our next step is to implement the accretion disk + reprocessing layer (AD+RL) model ~\citep{TiDE,Guillochon+2014ApJ} into our fitting code. In this model, an slim accretion disk is the source of UV and X-ray photons. During the super-Eddington phase, the wind launched from the accretion disk serves as the reprocessing layer that converts the high energy photons into optical photons \citep{SQ2009MNRAS}. In the late stage of the mass fallback, the accretion rate falls below the Eddington accretion rate, the disk wind is ceased and a bare accretion disk gradually exposed to the observer, which is responsible for the late-time plateau. The AD+RL model could also provide the ability of simultaneously fitting the X-ray and optical light curve, see for example the SED generated by \cite{Guillochon+2014ApJ} (their figure 9).

Before finishing this section, we briefly discuss the caveat of using equation~\ref{eq:compositeMag} to obtain the composite light curve. This equation assumes the flares of the two consecutive PTDEs evolves independently. However, this assumption might be inappropriate for two reasons. First, a gas disk could be leftover by the first PTDE (e.g. AT 2022dbl), which could interact with the debris of the second PTDE. Second, even if there is no residual gas disk, the debris from the first PTDE is still falling back by the time of the second PTDE (see Figure~\ref{fig:FallbackRate_b}). Since the two falling back debris lie in the same orbital plane, the interaction between them should also be accounted for. Although the light curves of the two TDEs presented here can be reproduced with equation~\ref{eq:compositeMag}, a comprehensive model is still needed. This should be done with the help of hydrodynamical simulation, which is beyond the scope of this work and we defer it to future study.

\begin{table}[htbp]
\caption{}
\label{table-params2}
\begin{center}
\begin{tabular}{c|cc|cc}
  \hline
  IAU name & $t_{\rm min,0}/\rm{day}$ & $\log t_{\nu}/\rm{day}$ & $\eta_{\rm fit}$ & $\eta_{\rm SD}$  \\
  \hline
  AT 2022dbl  & $21.96^{+0.48}_{-0.42}$ & $0.71^{+0.02}_{-0.03}$ & $1.82\times10^{-2}$ & $5.78\times10^{-3}$\\
  AT 2023adr  & $23.35^{+1.66}_{-1.75}$ & $-0.19^{+0.52}_{-0.54}$ & $7.4\times10^{-3}$ & $3.59\times10^{-2}$ \\
  \hline
\end{tabular}
\end{center}
\textbf{NOTES:} The first column gives the target name. The second column gives the orbital period of the most bound debris produced by the first PTDE, $t_{\rm min,0}$, which is also the evolution timescale of the mass fallback rate. The third column is the fitting result of viscous timescale. The fourth column is the median value of the fitted radiation efficiency. The fifth column is the median value of the radiation efficiency of the stream-disk collision (see Appendix~\ref{Appendix-eta_SD} for the details). We use the 16th and 84th percentiles of the posterior distribution to indicate the $1\sigma$ uncertainty in $t_{\rm min,0}$ and $\log{t_{\rm visc}}$.
\end{table}


\section{Summary}
\label{SECT:Summary}

In this work we develop a light curve model for the re-brightening TDEs, based on the scenario that the two consecutive flares are made by repeating PTDEs of the same star. Compared to the existing light curve fitting codes, we have the following updates in the modeling:

\begin{enumerate}
    \item The star capable of producing repeated PTDEs moves on an eccentric orbit around the central MBH. The mass fallback rate from an eccentric disruption is obtained by shifting the debris enengy distribution $dm/d\epsilon$ of parabolic disruption, towards the negative $\epsilon$ direction (equation~\ref{eq:Shift_of_dmde}).
    \item We calculated the variation of $\epsilon_{\rm orb}$ after every PTDE based on the formulae of \cite{ChenDaiLiu2024arXiv}. Since $\epsilon_{\rm orb}$ is an important quantity that determines both the time and the mass fallback rate of the next disruption, this step is necessary in accurately modeling the light curve.
    \item We fit the two flares in the re-brightening TDEs simultaneously, while the existing code could only handle one flare. We note that during the second flare, there is still some contribution from the first flare (Fig.~\ref{fig:FallbackRate_b}). Our approach (see equation~\ref{eq:compositeMag}) could model the second flare more accurately.
\end{enumerate}

As a demonstration, the light curves of two TDEs that have re-brightening feature (AT 2022dbl and AT 2023adr) are well reproduced by our fitting code. From these fittings, we obtained the physical parameters of our interests ($M_{\rm BH}$, $m_*$, $\beta$, $e$) for these two targets.

Our model is primarily designed for the re-brightening TDEs, it can also be applied to the single-flare TDEs. The star that give rise to single-flare TDE may have a different origin than the re-brightening TDE. As mentioned in Section~\ref{SECT:Introduction}, in almost all the previous studies, the orbital eccentricity $e$ of the disrupted star is fixed at unity. \cite{HZL2018} have tested this assumption using direct $N$-body simulations, and found that only a tiny fraction of the TDEs having $e=1$, while most of TDEs are marginally eccentric or marginally hyperbolic (see their Table 2). Although the deviation of $e$ from unity is very small (typically order of $10^{-3}$), it could already causes significant deviation in the mass fallback rate due to the large mass ratio $q$ between the MBH and the disrupted star (see equation~\ref{eq:E_orb(e)}). Hence, one should consider the non-parabolic disruptions when modeling the TDE light curves, and our fitting code is already capable of handling the non-parabolic disruptions. Besides, by analyzing the light curve of TDEs that only show single-flare up to date, we could hunt for the candidates that are capable of producing a second flare, which is an ongoing work in our group. For these candidates, we could also predict the time and brightness of the next flare (like we did in Section~\ref{SUBSECT:forecast}), so that the observers could get prepared in advance.

\cite{ZHL2023ApJ} have analytically derived the $\beta$ and $e$ distribution of TDEs occurred in a spherical galactic nuclei. They found that the TDEs are located in a restricted region in the $e$-$\beta$ parameter space (see Fig. 8 of \cite{ZHL2023ApJ}), whose boundary is determined by the density and velocity dispersion profile of the galactic nuclei. Therefore, extracting the information of $\beta$ and $e$ from the TDE light curve and comparing with the theoretical expectations, could provide constraint on the dynamical state of the TDE host galactic nuclei.

\pagebreak

\begin{acknowledgments}
The author thanks the anonymous referee for careful reading and providing constructive comments, which helps a lot in improving the paper.
The author acknowledges support from the ``Yunnan Provincial Key Laboratory of Survey Science" with project No. 202449CE340002. The author acknowledges supports from the “Science \& Technology Champion Project” (202005AB160002) and from two “Team Projects” – the “Innovation Team” (202105AE160021) and the “Top Team” (202305AT350002), all funded by the “Yunnan Revitalization Talent Support Program”. 
S.Z. thanks Ning Jiang and Zheyu Lin for useful discussions, and for kindly offering the photometric data of AT 2022dbl. S.Z. also thanks Hongwei Ge for useful comments and suggestions.
\end{acknowledgments}


\bibliography{RTDEs}

\begin{thebibliography}{}
\expandafter\ifx\csname natexlab\endcsname\relax\def\natexlab#1{#1}\fi
\providecommand{\url}[1]{\href{#1}{#1}}
\providecommand{\dodoi}[1]{doi:~\href{http://doi.org/#1}{\nolinkurl{#1}}}
\providecommand{\doeprint}[1]{\href{http://ascl.net/#1}{\nolinkurl{http://ascl.net/#1}}}
\providecommand{\doarXiv}[1]{\href{https://arxiv.org/abs/#1}{\nolinkurl{https://arxiv.org/abs/#1}}}

\bibitem[{{Arcodia} {et~al.}(2021){Arcodia}, {Merloni}, {Nandra}, {Buchner}, {Salvato}, {Pasham}, {Remillard}, {Comparat}, {Lamer}, {Ponti}, {Malyali}, {Wolf}, {Arzoumanian}, {Bogensberger}, {Buckley}, {Gendreau}, {Gromadzki}, {Kara}, {Krumpe}, {Markwardt}, {Ramos-Ceja}, {Rau}, {Schramm}, \& {Schwope}}]{Arcodia+2021}
{Arcodia}, R., {Merloni}, A., {Nandra}, K., {et~al.} 2021, \nat, 592, 704, \dodoi{10.1038/s41586-021-03394-6}

\bibitem[{{Bahcall} \& {Wolf}(1976)}]{BW1976}
{Bahcall}, J.~N., \& {Wolf}, R.~A. 1976, \apj, 209, 214, \dodoi{10.1086/154711}

\bibitem[{{Bandopadhyay} {et~al.}(2024){Bandopadhyay}, {Coughlin}, {Nixon}, \& {Pasham}}]{Bandopadhyay+2024ApJ}
{Bandopadhyay}, A., {Coughlin}, E.~R., {Nixon}, C.~J., \& {Pasham}, D.~R. 2024, \apj, 974, 80, \dodoi{10.3847/1538-4357/ad6a5a}

\bibitem[{{Bellm} {et~al.}(2019){Bellm}, {Kulkarni}, {Graham}, {Dekany}, {Smith}, {Riddle}, {Masci}, {Helou}, {Prince}, {Adams}, {Barbarino}, {Barlow}, {Bauer}, {Beck}, {Belicki}, {Biswas}, {Blagorodnova}, {Bodewits}, {Bolin}, {Brinnel}, {Brooke}, {Bue}, {Bulla}, {Burruss}, {Cenko}, {Chang}, {Connolly}, {Coughlin}, {Cromer}, {Cunningham}, {De}, {Delacroix}, {Desai}, {Duev}, {Eadie}, {Farnham}, {Feeney}, {Feindt}, {Flynn}, {Franckowiak}, {Frederick}, {Fremling}, {Gal-Yam}, {Gezari}, {Giomi}, {Goldstein}, {Golkhou}, {Goobar}, {Groom}, {Hacopians}, {Hale}, {Henning}, {Ho}, {Hover}, {Howell}, {Hung}, {Huppenkothen}, {Imel}, {Ip}, {Ivezi{\'c}}, {Jackson}, {Jones}, {Juric}, {Kasliwal}, {Kaspi}, {Kaye}, {Kelley}, {Kowalski}, {Kramer}, {Kupfer}, {Landry}, {Laher}, {Lee}, {Lin}, {Lin}, {Lunnan}, {Giomi}, {Mahabal}, {Mao}, {Miller}, {Monkewitz}, {Murphy}, {Ngeow}, {Nordin}, {Nugent}, {Ofek}, {Patterson}, {Penprase}, {Porter}, {Rauch}, {Rebbapragada}, {Reiley}, {Rigault}, {Rodriguez}, {van Roestel}, {Rusholme}, {van
  Santen}, {Schulze}, {Shupe}, {Singer}, {Soumagnac}, {Stein}, {Surace}, {Sollerman}, {Szkody}, {Taddia}, {Terek}, {Van Sistine}, {van Velzen}, {Vestrand}, {Walters}, {Ward}, {Ye}, {Yu}, {Yan}, \& {Zolkower}}]{Bellm2019PASP..131a8002B}
{Bellm}, E.~C., {Kulkarni}, S.~R., {Graham}, M.~J., {et~al.} 2019, \pasp, 131, 018002, \dodoi{10.1088/1538-3873/aaecbe}

\bibitem[{{Bonnerot} \& {Lu}(2022)}]{Bonnerot+Lu2022MNRAS}
{Bonnerot}, C., \& {Lu}, W. 2022, \mnras, 511, 2147, \dodoi{10.1093/mnras/stac146}

\bibitem[{{Cannizzo} {et~al.}(1990){Cannizzo}, {Lee}, \& {Goodman}}]{CLG1990}
{Cannizzo}, J.~K., {Lee}, H.~M., \& {Goodman}, J. 1990, \apj, 351, 38, \dodoi{10.1086/168442}

\bibitem[{{Charalampopoulos} {et~al.}(2023){Charalampopoulos}, {Pursiainen}, {Leloudas}, {Arcavi}, {Newsome}, {Schulze}, {Burke}, \& {Nicholl}}]{CPLA2023A&A}
{Charalampopoulos}, P., {Pursiainen}, M., {Leloudas}, G., {et~al.} 2023, \aap, 673, A95, \dodoi{10.1051/0004-6361/202245065}

\bibitem[{{Chen} {et~al.}(2024){Chen}, {Dai}, {Liu}, \& {Ou}}]{ChenDaiLiu2024arXiv}
{Chen}, J.-H., {Dai}, L., {Liu}, S.-F., \& {Ou}, J.-W. 2024, arXiv e-prints, arXiv:2408.10925, \dodoi{10.48550/arXiv.2408.10925}

\bibitem[{{Chen} {et~al.}(2022){Chen}, {Dou}, \& {Shen}}]{ChenDouShen2022ApJ}
{Chen}, J.-H., {Dou}, L.-M., \& {Shen}, R.-F. 2022, \apj, 928, 63, \dodoi{10.3847/1538-4357/ac558d}

\bibitem[{{Cohn} \& {Kulsrud}(1978)}]{CK1978}
{Cohn}, H., \& {Kulsrud}, R.~M. 1978, \apj, 226, 1087, \dodoi{10.1086/156685}

\bibitem[{{Cufari} {et~al.}(2022){Cufari}, {Coughlin}, \& {Nixon}}]{Cufari+2022ApJ}
{Cufari}, M., {Coughlin}, E.~R., \& {Nixon}, C.~J. 2022, \apjl, 929, L20, \dodoi{10.3847/2041-8213/ac6021}

\bibitem[{{Dai} {et~al.}(2013){Dai}, {Blandford}, \& {Eggleton}}]{DBE2013}
{Dai}, L., {Blandford}, R.~D., \& {Eggleton}, P.~P. 2013, \mnras, 434, 2940, \dodoi{10.1093/mnras/stt1208}

\bibitem[{{Dai} {et~al.}(2015){Dai}, {McKinney}, \& {Miller}}]{Dai+2015ApJ}
{Dai}, L., {McKinney}, J.~C., \& {Miller}, M.~C. 2015, \apjl, 812, L39, \dodoi{10.1088/2041-8205/812/2/L39}

\bibitem[{{Evans} {et~al.}(2023){Evans}, {Nixon}, {Campana}, {Charalampopoulos}, {Perley}, {Breeveld}, {Page}, {Oates}, {Eyles-Ferris}, {Malesani}, {Izzo}, {Goad}, {O'Brien}, {Osborne}, \& {Sbarufatti}}]{Evans+2023NatAs}
{Evans}, P.~A., {Nixon}, C.~J., {Campana}, S., {et~al.} 2023, Nature Astronomy, 7, 1368, \dodoi{10.1038/s41550-023-02073-y}

\bibitem[{{Faris} {et~al.}(2024){Faris}, {Arcavi}, {Makrygianni}, {Hiramatsu}, {Terreran}, {Farah}, {Howell}, {McCully}, {Newsome}, {Padilla Gonzalez}, {Pellegrino}, {Bostroem}, {Abojanb}, {Lam}, {Tomasella}, {Brink}, {Filippenko}, {French}, {Clark}, {Graur}, {Leloudas}, {Gromadzki}, {Anderson}, {Nicholl}, {Guti{\'e}rrez}, {Kankare}, {Inserra}, {Galbany}, {Reynolds}, {Mattila}, {Heikkil{\"a}}, {Wang}, {Onori}, {Wevers}, {Coughlin}, {Charalampopoulos}, \& {Johansson}}]{FAMH2024ApJ}
{Faris}, S., {Arcavi}, I., {Makrygianni}, L., {et~al.} 2024, \apj, 969, 104, \dodoi{10.3847/1538-4357/ad4a72}

\bibitem[{{Foreman-Mackey} {et~al.}(2013){Foreman-Mackey}, {Hogg}, {Lang}, \& {Goodman}}]{emcee2013PASP}
{Foreman-Mackey}, D., {Hogg}, D.~W., {Lang}, D., \& {Goodman}, J. 2013, \pasp, 125, 306, \dodoi{10.1086/670067}

\bibitem[{{Frank} \& {Rees}(1976)}]{FR1976}
{Frank}, J., \& {Rees}, M.~J. 1976, \mnras, 176, 633

\bibitem[{{Gafton} {et~al.}(2015){Gafton}, {Tejeda}, {Guillochon}, {Korobkin}, \& {Rosswog}}]{GTG2015}
{Gafton}, E., {Tejeda}, E., {Guillochon}, J., {Korobkin}, O., \& {Rosswog}, S. 2015, \mnras, 449, 771, \dodoi{10.1093/mnras/stv350}

\bibitem[{{Ge} {et~al.}(2010){Ge}, {Hjellming}, {Webbink}, {Chen}, \& {Han}}]{GHW2010ApJ}
{Ge}, H., {Hjellming}, M.~S., {Webbink}, R.~F., {Chen}, X., \& {Han}, Z. 2010, \apj, 717, 724, \dodoi{10.1088/0004-637X/717/2/724}

\bibitem[{{Ge} {et~al.}(2015){Ge}, {Webbink}, {Chen}, \& {Han}}]{Ge+2015ApJ}
{Ge}, H., {Webbink}, R.~F., {Chen}, X., \& {Han}, Z. 2015, \apj, 812, 40, \dodoi{10.1088/0004-637X/812/1/40}

\bibitem[{{Ge} {et~al.}(2020){Ge}, {Webbink}, {Chen}, \& {Han}}]{GWC2020ApJ}
---. 2020, \apj, 899, 132, \dodoi{10.3847/1538-4357/aba7b7}

\bibitem[{{Ge} {et~al.}(2024){Ge}, {Tout}, {Chen}, {Wang}, {Xiong}, {Zhang}, {Li}, {Liu}, \& {Han}}]{GTC2024}
{Ge}, H., {Tout}, C.~A., {Chen}, X., {et~al.} 2024, \apj, 975, 254, \dodoi{10.3847/1538-4357/ad7ea6}

\bibitem[{{Gezari}(2021)}]{Gezari2021ARAA}
{Gezari}, S. 2021, \araa, 59, 21, \dodoi{10.1146/annurev-astro-111720-030029}

\bibitem[{{Giustini} {et~al.}(2020){Giustini}, {Miniutti}, \& {Saxton}}]{GMS2020}
{Giustini}, M., {Miniutti}, G., \& {Saxton}, R.~D. 2020, \aap, 636, L2, \dodoi{10.1051/0004-6361/202037610}

\bibitem[{{Golightly} {et~al.}(2019){Golightly}, {Coughlin}, \& {Nixon}}]{Golightly+2019ApJ}
{Golightly}, E. C.~A., {Coughlin}, E.~R., \& {Nixon}, C.~J. 2019, \apj, 872, 163, \dodoi{10.3847/1538-4357/aafd2f}

\bibitem[{{Gomez} {et~al.}(2020){Gomez}, {Nicholl}, {Short}, {Margutti}, {Alexander}, {Blanchard}, {Berger}, {Eftekhari}, {Schulze}, {Anderson}, {Arcavi}, {Chornock}, {Cowperthwaite}, {Galbany}, {Herzog}, {Hiramatsu}, {Hosseinzadeh}, {Laskar}, {M{\"u}ller Bravo}, {Patton}, \& {Terreran}}]{Gomez+2020}
{Gomez}, S., {Nicholl}, M., {Short}, P., {et~al.} 2020, \mnras, 497, 1925, \dodoi{10.1093/mnras/staa2099}

\bibitem[{{Graham} {et~al.}(2019){Graham}, {Kulkarni}, {Bellm}, {Adams}, {Barbarino}, {Blagorodnova}, {Bodewits}, {Bolin}, {Brady}, {Cenko}, {Chang}, {Coughlin}, {De}, {Eadie}, {Farnham}, {Feindt}, {Franckowiak}, {Fremling}, {Gezari}, {Ghosh}, {Goldstein}, {Golkhou}, {Goobar}, {Ho}, {Huppenkothen}, {Ivezi{\'c}}, {Jones}, {Juric}, {Kaplan}, {Kasliwal}, {Kelley}, {Kupfer}, {Lee}, {Lin}, {Lunnan}, {Mahabal}, {Miller}, {Ngeow}, {Nugent}, {Ofek}, {Prince}, {Rauch}, {van Roestel}, {Schulze}, {Singer}, {Sollerman}, {Taddia}, {Yan}, {Ye}, {Yu}, {Barlow}, {Bauer}, {Beck}, {Belicki}, {Biswas}, {Brinnel}, {Brooke}, {Bue}, {Bulla}, {Burruss}, {Connolly}, {Cromer}, {Cunningham}, {Dekany}, {Delacroix}, {Desai}, {Duev}, {Feeney}, {Flynn}, {Frederick}, {Gal-Yam}, {Giomi}, {Groom}, {Hacopians}, {Hale}, {Helou}, {Henning}, {Hover}, {Hillenbrand}, {Howell}, {Hung}, {Imel}, {Ip}, {Jackson}, {Kaspi}, {Kaye}, {Kowalski}, {Kramer}, {Kuhn}, {Landry}, {Laher}, {Mao}, {Masci}, {Monkewitz}, {Murphy}, {Nordin}, {Patterson}, {Penprase},
  {Porter}, {Rebbapragada}, {Reiley}, {Riddle}, {Rigault}, {Rodriguez}, {Rusholme}, {van Santen}, {Shupe}, {Smith}, {Soumagnac}, {Stein}, {Surace}, {Szkody}, {Terek}, {Van Sistine}, {van Velzen}, {Vestrand}, {Walters}, {Ward}, {Zhang}, \& {Zolkower}}]{Graham2019PASP}
{Graham}, M.~J., {Kulkarni}, S.~R., {Bellm}, E.~C., {et~al.} 2019, \pasp, 131, 078001, \dodoi{10.1088/1538-3873/ab006c}

\bibitem[{{Guillochon} {et~al.}(2014){Guillochon}, {Manukian}, \& {Ramirez-Ruiz}}]{Guillochon+2014ApJ}
{Guillochon}, J., {Manukian}, H., \& {Ramirez-Ruiz}, E. 2014, \apj, 783, 23, \dodoi{10.1088/0004-637X/783/1/23}

\bibitem[{{Guillochon} {et~al.}(2018){Guillochon}, {Nicholl}, {Villar}, {Mockler}, {Narayan}, {Mandel}, {Berger}, \& {Williams}}]{MOSFiT}
{Guillochon}, J., {Nicholl}, M., {Villar}, V.~A., {et~al.} 2018, \apjs, 236, 6, \dodoi{10.3847/1538-4365/aab761}

\bibitem[{{Guillochon} \& {Ramirez-Ruiz}(2013)}]{GRR2013}
{Guillochon}, J., \& {Ramirez-Ruiz}, E. 2013, \apj, 767, 25, \dodoi{10.1088/0004-637X/767/1/25}

\bibitem[{{Guolo} {et~al.}(2024){Guolo}, {Pasham}, {Zaja{\v{c}}ek}, {Coughlin}, {Gezari}, {Sukov{\'a}}, {Wevers}, {Witzany}, {Tombesi}, {van Velzen}, {Alexander}, {Yao}, {Arcodia}, {Karas}, {Miller-Jones}, {Remillard}, {Gendreau}, \& {Ferrara}}]{Guolo+2024NatAs}
{Guolo}, M., {Pasham}, D.~R., {Zaja{\v{c}}ek}, M., {et~al.} 2024, Nature Astronomy, 8, 347, \dodoi{10.1038/s41550-023-02178-4}

\bibitem[{{Hammerstein} {et~al.}(2023){Hammerstein}, {van Velzen}, {Gezari}, {Cenko}, {Yao}, {Ward}, {Frederick}, {Villanueva}, {Somalwar}, {Graham}, {Kulkarni}, {Stern}, {Andreoni}, {Bellm}, {Dekany}, {Dhawan}, {Drake}, {Fremling}, {Gatkine}, {Groom}, {Ho}, {Kasliwal}, {Karambelkar}, {Kool}, {Masci}, {Medford}, {Perley}, {Purdum}, {van Roestel}, {Sharma}, {Sollerman}, {Taggart}, \& {Yan}}]{HvVC2023}
{Hammerstein}, E., {van Velzen}, S., {Gezari}, S., {et~al.} 2023, \apj, 942, 9, \dodoi{10.3847/1538-4357/aca283}

\bibitem[{{Hampel} {et~al.}(2022){Hampel}, {Komossa}, {Greiner}, {Reiprich}, {Freyberg}, \& {Erben}}]{HKG2022}
{Hampel}, J., {Komossa}, S., {Greiner}, J., {et~al.} 2022, Research in Astronomy and Astrophysics, 22, 055004, \dodoi{10.1088/1674-4527/ac5800}

\bibitem[{{Hayasaki} {et~al.}(2013){Hayasaki}, {Stone}, \& {Loeb}}]{HSL2013MNRAS}
{Hayasaki}, K., {Stone}, N., \& {Loeb}, A. 2013, \mnras, 434, 909, \dodoi{10.1093/mnras/stt871}

\bibitem[{{Hayasaki} {et~al.}(2018){Hayasaki}, {Zhong}, {Li}, {Berczik}, \& {Spurzem}}]{HZL2018}
{Hayasaki}, K., {Zhong}, S., {Li}, S., {Berczik}, P., \& {Spurzem}, R. 2018, \apj, 855, 129, \dodoi{10.3847/1538-4357/aab0a5}

\bibitem[{{Hills}(1988)}]{Hills1988}
{Hills}, J.~G. 1988, \nat, 331, 687, \dodoi{10.1038/331687a0}

\bibitem[{{Hjellming} \& {Webbink}(1987)}]{HW1987}
{Hjellming}, M.~S., \& {Webbink}, R.~F. 1987, \apj, 318, 794, \dodoi{10.1086/165412}

\bibitem[{{Huang} {et~al.}(2024{\natexlab{a}}){Huang}, {Jiang}, {Zhu}, {Wang}, {Wang}, {Wang}, {Gan}, {Liang}, {Qin}, {Lin}, {Xu}, {Cai}, {Jiang}, {Kong}, {Li}, {li}, {Wang}, {Xu}, {Xue}, {Yuan}, {Cheng}, {Fan}, {Gao}, {Hu}, {Hu}, {Li}, {Li}, {Liang}, {Liu}, {Liu}, {Lou}, {Luo}, {Qian}, {Tang}, {Wan}, {Wang}, {Wang}, {Yang}, {Yao}, {Zhang}, {Zhang}, {Zhao}, {Zheng}, {Zhu}, \& {Zuo}}]{HuangJiangZhu2024ApJ}
{Huang}, S., {Jiang}, N., {Zhu}, J., {et~al.} 2024{\natexlab{a}}, \apjl, 964, L22, \dodoi{10.3847/2041-8213/ad319f}

\bibitem[{{Huang} {et~al.}(2024{\natexlab{b}}){Huang}, {Davis}, \& {Jiang}}]{Huang+2024ApJ}
{Huang}, X., {Davis}, S.~W., \& {Jiang}, Y.-f. 2024{\natexlab{b}}, \apj, 974, 165, \dodoi{10.3847/1538-4357/ad6c39}

\bibitem[{{Ivezi{\'c}} {et~al.}(2019){Ivezi{\'c}}, {Kahn}, {Tyson}, {Abel}, {Acosta}, {Allsman}, {Alonso}, {AlSayyad}, {Anderson}, {Andrew}, {Angel}, {Angeli}, {Ansari}, {Antilogus}, {Araujo}, {Armstrong}, {Arndt}, {Astier}, {Aubourg}, {Auza}, {Axelrod}, {Bard}, {Barr}, {Barrau}, {Bartlett}, {Bauer}, {Bauman}, {Baumont}, {Bechtol}, {Bechtol}, {Becker}, {Becla}, {Beldica}, {Bellavia}, {Bianco}, {Biswas}, {Blanc}, {Blazek}, {Blandford}, {Bloom}, {Bogart}, {Bond}, {Booth}, {Borgland}, {Borne}, {Bosch}, {Boutigny}, {Brackett}, {Bradshaw}, {Brandt}, {Brown}, {Bullock}, {Burchat}, {Burke}, {Cagnoli}, {Calabrese}, {Callahan}, {Callen}, {Carlin}, {Carlson}, {Chandrasekharan}, {Charles-Emerson}, {Chesley}, {Cheu}, {Chiang}, {Chiang}, {Chirino}, {Chow}, {Ciardi}, {Claver}, {Cohen-Tanugi}, {Cockrum}, {Coles}, {Connolly}, {Cook}, {Cooray}, {Covey}, {Cribbs}, {Cui}, {Cutri}, {Daly}, {Daniel}, {Daruich}, {Daubard}, {Daues}, {Dawson}, {Delgado}, {Dellapenna}, {de Peyster}, {de Val-Borro}, {Digel}, {Doherty}, {Dubois},
  {Dubois-Felsmann}, {Durech}, {Economou}, {Eifler}, {Eracleous}, {Emmons}, {Fausti Neto}, {Ferguson}, {Figueroa}, {Fisher-Levine}, {Focke}, {Foss}, {Frank}, {Freemon}, {Gangler}, {Gawiser}, {Geary}, {Gee}, {Geha}, {Gessner}, {Gibson}, {Gilmore}, {Glanzman}, {Glick}, {Goldina}, {Goldstein}, {Goodenow}, {Graham}, {Gressler}, {Gris}, {Guy}, {Guyonnet}, {Haller}, {Harris}, {Hascall}, {Haupt}, {Hernandez}, {Herrmann}, {Hileman}, {Hoblitt}, {Hodgson}, {Hogan}, {Howard}, {Huang}, {Huffer}, {Ingraham}, {Innes}, {Jacoby}, {Jain}, {Jammes}, {Jee}, {Jenness}, {Jernigan}, {Jevremovi{\'c}}, {Johns}, {Johnson}, {Johnson}, {Jones}, {Juramy-Gilles}, {Juri{\'c}}, {Kalirai}, {Kallivayalil}, {Kalmbach}, {Kantor}, {Karst}, {Kasliwal}, {Kelly}, {Kessler}, {Kinnison}, {Kirkby}, {Knox}, {Kotov}, {Krabbendam}, {Krughoff}, {Kub{\'a}nek}, {Kuczewski}, {Kulkarni}, {Ku}, {Kurita}, {Lage}, {Lambert}, {Lange}, {Langton}, {Le Guillou}, {Levine}, {Liang}, {Lim}, {Lintott}, {Long}, {Lopez}, {Lotz}, {Lupton}, {Lust}, {MacArthur}, {Mahabal},
  {Mandelbaum}, {Markiewicz}, {Marsh}, {Marshall}, {Marshall}, {May}, {McKercher}, {McQueen}, {Meyers}, {Migliore}, {Miller}, {Mills}, {Miraval}, {Moeyens}, {Moolekamp}, {Monet}, {Moniez}, {Monkewitz}, {Montgomery}, {Morrison}, {Mueller}, {Muller}, {Mu{\~n}oz Arancibia}, {Neill}, {Newbry}, {Nief}, {Nomerotski}, {Nordby}, {O'Connor}, {Oliver}, {Olivier}, {Olsen}, {O'Mullane}, {Ortiz}, {Osier}, {Owen}, {Pain}, {Palecek}, {Parejko}, {Parsons}, {Pease}, {Peterson}, {Peterson}, {Petravick}, {Libby Petrick}, {Petry}, {Pierfederici}, {Pietrowicz}, {Pike}, {Pinto}, {Plante}, {Plate}, {Plutchak}, {Price}, {Prouza}, {Radeka}, {Rajagopal}, {Rasmussen}, {Regnault}, {Reil}, {Reiss}, {Reuter}, {Ridgway}, {Riot}, {Ritz}, {Robinson}, {Roby}, {Roodman}, {Rosing}, {Roucelle}, {Rumore}, {Russo}, {Saha}, {Sassolas}, {Schalk}, {Schellart}, {Schindler}, {Schmidt}, {Schneider}, {Schneider}, {Schoening}, {Schumacher}, {Schwamb}, {Sebag}, {Selvy}, {Sembroski}, {Seppala}, {Serio}, {Serrano}, {Shaw}, {Shipsey}, {Sick}, {Silvestri},
  {Slater}, {Smith}, {Smith}, {Sobhani}, {Soldahl}, {Storrie-Lombardi}, {Stover}, {Strauss}, {Street}, {Stubbs}, {Sullivan}, {Sweeney}, {Swinbank}, {Szalay}, {Takacs}, {Tether}, {Thaler}, {Thayer}, {Thomas}, {Thornton}, {Thukral}, {Tice}, {Trilling}, {Turri}, {Van Berg}, {Vanden Berk}, {Vetter}, {Virieux}, {Vucina}, {Wahl}, {Walkowicz}, {Walsh}, {Walter}, {Wang}, {Wang}, {Warner}, {Wiecha}, {Willman}, {Winters}, {Wittman}, {Wolff}, {Wood-Vasey}, {Wu}, {Xin}, {Yoachim}, \& {Zhan}}]{Ivezic2019ApJ...873..111I}
{Ivezi{\'c}}, {\v{Z}}., {Kahn}, S.~M., {Tyson}, J.~A., {et~al.} 2019, \apj, 873, 111, \dodoi{10.3847/1538-4357/ab042c}

\bibitem[{{Jiang} {et~al.}(2019){Jiang}, {Wang}, {Mou}, {Liu}, {Dou}, {Sheng}, \& {Wang}}]{JWM2019}
{Jiang}, N., {Wang}, T., {Mou}, G., {et~al.} 2019, \apj, 871, 15, \dodoi{10.3847/1538-4357/aaf6b2}

\bibitem[{{Jiang} {et~al.}(2016){Jiang}, {Guillochon}, \& {Loeb}}]{Jiang_2016ApJ}
{Jiang}, Y.-F., {Guillochon}, J., \& {Loeb}, A. 2016, \apj, 830, 125, \dodoi{10.3847/0004-637X/830/2/125}

\bibitem[{{Komossa} \& {Bade}(1999)}]{KB1999A&A}
{Komossa}, S., \& {Bade}, N. 1999, \aap, 343, 775, \dodoi{10.48550/arXiv.astro-ph/9901141}

\bibitem[{{Kormendy} \& {Ho}(2013)}]{KormendyHo2013ARAA}
{Kormendy}, J., \& {Ho}, L.~C. 2013, \araa, 51, 511, \dodoi{10.1146/annurev-astro-082708-101811}

\bibitem[{{Kov{\'a}cs-Stermeczky} \& {Vink{\'o}}(2023{\natexlab{a}})}]{TiDE}
{Kov{\'a}cs-Stermeczky}, Z.~V., \& {Vink{\'o}}, J. 2023{\natexlab{a}}, \pasp, 135, 034102, \dodoi{10.1088/1538-3873/acb9bb}

\bibitem[{{Kov{\'a}cs-Stermeczky} \& {Vink{\'o}}(2023{\natexlab{b}})}]{KV2023}
---. 2023{\natexlab{b}}, \pasp, 135, 104102, \dodoi{10.1088/1538-3873/acf8f8}

\bibitem[{{Kumar} \& {Goodman}(1996)}]{KumarGoodman1996ApJ}
{Kumar}, P., \& {Goodman}, J. 1996, \apj, 466, 946, \dodoi{10.1086/177565}

\bibitem[{{Lin} {et~al.}(2024){Lin}, {Jiang}, {Wang}, {Kong}, {Li}, {He}, {Wang}, {Zhu}, {Li}, {Jiang}, {Singh}, {Teja}, {Sahu}, {Jin}, {Maeda}, \& {Huang}}]{LinJiangWang+2024ApJ}
{Lin}, Z., {Jiang}, N., {Wang}, T., {et~al.} 2024, \apjl, 971, L26, \dodoi{10.3847/2041-8213/ad638e}

\bibitem[{{Liu} {et~al.}(2023{\natexlab{a}}){Liu}, {Mockler}, {Ramirez-Ruiz}, {Yarza}, {Law-Smith}, {Naoz}, {Melchor}, \& {Rose}}]{LMRR2023ApJ}
{Liu}, C., {Mockler}, B., {Ramirez-Ruiz}, E., {et~al.} 2023{\natexlab{a}}, \apj, 944, 184, \dodoi{10.3847/1538-4357/acafe1}

\bibitem[{{Liu} {et~al.}(2024{\natexlab{a}}){Liu}, {Yarza}, \& {Ramirez-Ruiz}}]{Liu+2024arXiv}
{Liu}, C., {Yarza}, R., \& {Ramirez-Ruiz}, E. 2024{\natexlab{a}}, arXiv e-prints, arXiv:2406.01670, \dodoi{10.48550/arXiv.2406.01670}

\bibitem[{{Liu} {et~al.}(2023{\natexlab{b}}){Liu}, {Malyali}, {Krumpe}, {Homan}, {Goodwin}, {Grotova}, {Kawka}, {Rau}, {Merloni}, {Anderson}, {Miller-Jones}, {Markowitz}, {Ciroi}, {Di Mille}, {Schramm}, {Tang}, {Buckley}, {Gromadzki}, {Jin}, \& {Buchner}}]{LMK2023}
{Liu}, Z., {Malyali}, A., {Krumpe}, M., {et~al.} 2023{\natexlab{b}}, \aap, 669, A75, \dodoi{10.1051/0004-6361/202244805}

\bibitem[{{Liu} {et~al.}(2024{\natexlab{b}}){Liu}, {Ryu}, {Goodwin}, {Rau}, {Homan}, {Krumpe}, {Merloni}, {Grotova}, {Anderson}, {Malyali}, \& {Miller-Jones}}]{LRG2024}
{Liu}, Z., {Ryu}, T., {Goodwin}, A.~J., {et~al.} 2024{\natexlab{b}}, \aap, 683, L13, \dodoi{10.1051/0004-6361/202348682}

\bibitem[{{Llamas Lanza} {et~al.}(2024){Llamas Lanza}, {Quintin}, {Russeil}, {Ishida}, {Peloton}, {Karpov}, \& {Pruzhinskaya}}]{LQ2024TNSAN}
{Llamas Lanza}, M., {Quintin}, E., {Russeil}, E., {et~al.} 2024, Transient Name Server AstroNote, 178, 1

\bibitem[{{Malyali} {et~al.}(2023){Malyali}, {Liu}, {Rau}, {Grotova}, {Merloni}, {Goodwin}, {Anderson}, {Miller-Jones}, {Kawka}, {Arcodia}, {Buchner}, {Nandra}, {Homan}, \& {Krumpe}}]{MLR2023}
{Malyali}, A., {Liu}, Z., {Rau}, A., {et~al.} 2023, \mnras, 520, 3549, \dodoi{10.1093/mnras/stad022}

\bibitem[{{Mandel} \& {Levin}(2015)}]{Mandel_Levin2015ApJ}
{Mandel}, I., \& {Levin}, Y. 2015, \apjl, 805, L4, \dodoi{10.1088/2041-8205/805/1/L4}

\bibitem[{{Manukian} {et~al.}(2013){Manukian}, {Guillochon}, {Ramirez-Ruiz}, \& {O'Leary}}]{MGR2013}
{Manukian}, H., {Guillochon}, J., {Ramirez-Ruiz}, E., \& {O'Leary}, R.~M. 2013, \apjl, 771, L28, \dodoi{10.1088/2041-8205/771/2/L28}

\bibitem[{{Miniutti} {et~al.}(2019){Miniutti}, {Saxton}, {Giustini}, {Alexander}, {Fender}, {Heywood}, {Monageng}, {Coriat}, {Tzioumis}, {Read}, {Knigge}, {Gandhi}, {Pretorius}, \& {Ag{\'\i}s-Gonz{\'a}lez}}]{Miniutti+2019}
{Miniutti}, G., {Saxton}, R.~D., {Giustini}, M., {et~al.} 2019, \nat, 573, 381, \dodoi{10.1038/s41586-019-1556-x}

\bibitem[{{Mockler} {et~al.}(2019){Mockler}, {Guillochon}, \& {Ramirez-Ruiz}}]{MGR2019}
{Mockler}, B., {Guillochon}, J., \& {Ramirez-Ruiz}, E. 2019, \apj, 872, 151, \dodoi{10.3847/1538-4357/ab010f}

\bibitem[{{Mockler} \& {Ramirez-Ruiz}(2021)}]{MR2021ApJ}
{Mockler}, B., \& {Ramirez-Ruiz}, E. 2021, \apj, 906, 101, \dodoi{10.3847/1538-4357/abc955}

\bibitem[{{Mummery} \& {Balbus}(2020)}]{MB2020MNRAS}
{Mummery}, A., \& {Balbus}, S.~A. 2020, \mnras, 492, 5655, \dodoi{10.1093/mnras/staa192}

\bibitem[{{Mummery} {et~al.}(2024){Mummery}, {van Velzen}, {Nathan}, {Ingram}, {Hammerstein}, {Fraser-Taliente}, \& {Balbus}}]{MvVN2024MNRAS}
{Mummery}, A., {van Velzen}, S., {Nathan}, E., {et~al.} 2024, \mnras, 527, 2452, \dodoi{10.1093/mnras/stad3001}

\bibitem[{{Nicholl} {et~al.}(2020){Nicholl}, {Wevers}, {Oates}, {Alexander}, {Leloudas}, {Onori}, {Jerkstrand}, {Gomez}, {Campana}, {Arcavi}, {Charalampopoulos}, {Gromadzki}, {Ihanec}, {Jonker}, {Lawrence}, {Mandel}, {Schulze}, {Short}, {Burke}, {McCully}, {Hiramatsu}, {Howell}, {Pellegrino}, {Abbot}, {Anderson}, {Berger}, {Blanchard}, {Cannizzaro}, {Chen}, {Dennefeld}, {Galbany}, {Gonz{\'a}lez-Gait{\'a}n}, {Hosseinzadeh}, {Inserra}, {Irani}, {Kuin}, {M{\"u}ller-Bravo}, {Pineda}, {Ross}, {Roy}, {Smartt}, {Smith}, {Tucker}, {Wyrzykowski}, \& {Young}}]{Nicholl+2020MNRAS}
{Nicholl}, M., {Wevers}, T., {Oates}, S.~R., {et~al.} 2020, \mnras, 499, 482, \dodoi{10.1093/mnras/staa2824}

\bibitem[{{Park} \& {Hayasaki}(2020)}]{PH2020ApJ}
{Park}, G., \& {Hayasaki}, K. 2020, \apj, 900, 3, \dodoi{10.3847/1538-4357/ab9ebb}

\bibitem[{{Paxton} {et~al.}(2011){Paxton}, {Bildsten}, {Dotter}, {Herwig}, {Lesaffre}, \& {Timmes}}]{MESA}
{Paxton}, B., {Bildsten}, L., {Dotter}, A., {et~al.} 2011, \apjs, 192, 3, \dodoi{10.1088/0067-0049/192/1/3}

\bibitem[{{Payne} {et~al.}(2021){Payne}, {Shappee}, {Hinkle}, {Vallely}, {Kochanek}, {Holoien}, {Auchettl}, {Stanek}, {Thompson}, {Neustadt}, {Tucker}, {Armstrong}, {Brimacombe}, {Cacella}, {Cornect}, {Denneau}, {Fausnaugh}, {Flewelling}, {Grupe}, {Heinze}, {Lopez}, {Monard}, {Prieto}, {Schneider}, {Sheppard}, {Tonry}, \& {Weiland}}]{Payne+2021ApJ}
{Payne}, A.~V., {Shappee}, B.~J., {Hinkle}, J.~T., {et~al.} 2021, \apj, 910, 125, \dodoi{10.3847/1538-4357/abe38d}

\bibitem[{{Pfahl}(2005)}]{Pfahl2005ApJ}
{Pfahl}, E. 2005, \apj, 626, 849, \dodoi{10.1086/430167}

\bibitem[{{Piran} {et~al.}(2015){Piran}, {Svirski}, {Krolik}, {Cheng}, \& {Shiokawa}}]{Piran+2015ApJ}
{Piran}, T., {Svirski}, G., {Krolik}, J., {Cheng}, R.~M., \& {Shiokawa}, H. 2015, \apj, 806, 164, \dodoi{10.1088/0004-637X/806/2/164}

\bibitem[{{Podsiadlowski}(1996)}]{Podsiadlowski1996MNRAS}
{Podsiadlowski}, P. 1996, \mnras, 279, 1104, \dodoi{10.1093/mnras/279.4.1104}

\bibitem[{{Rees}(1988)}]{Rees1988}
{Rees}, M.~J. 1988, \nat, 333, 523, \dodoi{10.1038/333523a0}

\bibitem[{{Ryu} {et~al.}(2020){Ryu}, {Krolik}, {Piran}, \& {Noble}}]{Ryu+2020ApJ}
{Ryu}, T., {Krolik}, J., {Piran}, T., \& {Noble}, S.~C. 2020, \apj, 904, 100, \dodoi{10.3847/1538-4357/abb3ce}

\bibitem[{{Ryu} {et~al.}(2023){Ryu}, {Krolik}, {Piran}, {Noble}, \& {Avara}}]{Ryu+2023ApJ.957.12R}
{Ryu}, T., {Krolik}, J., {Piran}, T., {Noble}, S.~C., \& {Avara}, M. 2023, \apj, 957, 12, \dodoi{10.3847/1538-4357/acf5de}

\bibitem[{{Schlafly} \& {Finkbeiner}(2011)}]{SF2011ApJ}
{Schlafly}, E.~F., \& {Finkbeiner}, D.~P. 2011, \apj, 737, 103, \dodoi{10.1088/0004-637X/737/2/103}

\bibitem[{{Shakura} \& {Sunyaev}(1973)}]{SS1973}
{Shakura}, N.~I., \& {Sunyaev}, R.~A. 1973, \aap, 24, 337

\bibitem[{{Sharma} {et~al.}(2024){Sharma}, {Price}, \& {Heger}}]{SPH2024MNRAS}
{Sharma}, M., {Price}, D.~J., \& {Heger}, A. 2024, \mnras, 532, 89, \dodoi{10.1093/mnras/stae1455}

\bibitem[{{Shiokawa} {et~al.}(2015){Shiokawa}, {Krolik}, {Cheng}, {Piran}, \& {Noble}}]{Shiokawa+2015ApJ}
{Shiokawa}, H., {Krolik}, J.~H., {Cheng}, R.~M., {Piran}, T., \& {Noble}, S.~C. 2015, \apj, 804, 85, \dodoi{10.1088/0004-637X/804/2/85}

\bibitem[{{Somalwar} {et~al.}(2023){Somalwar}, {Ravi}, {Yao}, {Guolo}, {Graham}, {Hammerstein}, {Lu}, {Nicholl}, {Sharma}, {Stein}, {van Velzen}, {Bellm}, {Coughlin}, {Groom}, {Masci}, \& {Riddle}}]{Somalwar2023arXiv}
{Somalwar}, J.~J., {Ravi}, V., {Yao}, Y., {et~al.} 2023, arXiv e-prints, arXiv:2310.03782, \dodoi{10.48550/arXiv.2310.03782}

\bibitem[{{Steinberg} \& {Stone}(2024)}]{Steinberg+Stone2024Natur}
{Steinberg}, E., \& {Stone}, N.~C. 2024, \nat, 625, 463, \dodoi{10.1038/s41586-023-06875-y}

\bibitem[{{Strubbe} \& {Quataert}(2009)}]{SQ2009MNRAS}
{Strubbe}, L.~E., \& {Quataert}, E. 2009, \mnras, 400, 2070, \dodoi{10.1111/j.1365-2966.2009.15599.x}

\bibitem[{{Sun} {et~al.}(2024){Sun}, {Jiang}, {Dou}, {Shu}, {Zhu}, {Dong}, {Buckley}, {Cenko}, {Fan}, {Gromadzki}, {Liu}, {Wang}, {Wang}, {Wang}, {Wu}, {Yang}, {Zhang}, {Zhang}, \& {Zhang}}]{Sun+2024arXiv}
{Sun}, L., {Jiang}, N., {Dou}, L., {et~al.} 2024, arXiv e-prints, arXiv:2410.09720, \dodoi{10.48550/arXiv.2410.09720}

\bibitem[{{Tout} {et~al.}(1996){Tout}, {Pols}, {Eggleton}, \& {Han}}]{Tout+1996MNRAS}
{Tout}, C.~A., {Pols}, O.~R., {Eggleton}, P.~P., \& {Han}, Z. 1996, \mnras, 281, 257, \dodoi{10.1093/mnras/281.1.257}

\bibitem[{{van Velzen} {et~al.}(2019){van Velzen}, {Stone}, {Metzger}, {Gezari}, {Brown}, \& {Fruchter}}]{vVSM2019ApJ}
{van Velzen}, S., {Stone}, N.~C., {Metzger}, B.~D., {et~al.} 2019, \apj, 878, 82, \dodoi{10.3847/1538-4357/ab1844}

\bibitem[{{van Velzen} {et~al.}(2021){van Velzen}, {Gezari}, {Hammerstein}, {Roth}, {Frederick}, {Ward}, {Hung}, {Cenko}, {Stein}, {Perley}, {Taggart}, {Foley}, {Sollerman}, {Blagorodnova}, {Andreoni}, {Bellm}, {Brinnel}, {De}, {Dekany}, {Feeney}, {Fremling}, {Giomi}, {Golkhou}, {Graham}, {Ho}, {Kasliwal}, {Kilpatrick}, {Kulkarni}, {Kupfer}, {Laher}, {Mahabal}, {Masci}, {Miller}, {Nordin}, {Riddle}, {Rusholme}, {van Santen}, {Sharma}, {Shupe}, \& {Soumagnac}}]{vanVelzen+2021ApJ}
{van Velzen}, S., {Gezari}, S., {Hammerstein}, E., {et~al.} 2021, \apj, 908, 4, \dodoi{10.3847/1538-4357/abc258}

\bibitem[{{Veres} {et~al.}(2024){Veres}, {Franckowiak}, {van Velzen}, {Adebahr}, {Taziaux}, {Necker}, {Stein}, {Kier}, {Mueller}, {Bomans}, {Jordana-Mitjans}, {Kowalski}, {Hammerstein}, {Marci-Boehncke}, {Reusch}, {Garrappa}, {Rose}, \& {Kashyap Das}}]{Veres+2024arXiv}
{Veres}, P.~M., {Franckowiak}, A., {van Velzen}, S., {et~al.} 2024, arXiv e-prints, arXiv:2408.17419, \dodoi{10.48550/arXiv.2408.17419}

\bibitem[{{Wang} {et~al.}(2022){Wang}, {Yin}, {Ma}, \& {Wu}}]{WYM2022}
{Wang}, M., {Yin}, J., {Ma}, Y., \& {Wu}, Q. 2022, \apj, 933, 225, \dodoi{10.3847/1538-4357/ac75e6}

\bibitem[{{Wang} {et~al.}(2023){Wang}, {Liu}, {Cai}, {Geng}, {Fang}, {He}, {Jiang}, {Jiang}, {Kong}, {Li}, {Li}, {Luo}, {Pan}, {Wu}, {Yang}, {Yu}, {Zheng}, {Zhu}, {Cai}, {Chen}, {Chen}, {Dai}, {Fan}, {Fan}, {Fang}, {He}, {Hu}, {Hu}, {Jin}, {Jiang}, {Li}, {Li}, {Li}, {Liang}, {Lin}, {Liu}, {Liu}, {Liu}, {Liu}, {Liu}, {Lou}, {Qu}, {Sheng}, {Shi}, {Shu}, {Su}, {Sun}, {Wang}, {Wang}, {Wang}, {Wang}, {Wei}, {Wei}, {Xue}, {Yan}, {Yang}, {Yuan}, {Yuan}, {Zhang}, {Zhang}, {Zhao}, \& {Zhao}}]{Wang+2023SCPMA}
{Wang}, T., {Liu}, G., {Cai}, Z., {et~al.} 2023, Science China Physics, Mechanics, and Astronomy, 66, 109512, \dodoi{10.1007/s11433-023-2197-5}

\bibitem[{{Wang} {et~al.}(2024){Wang}, {Wang}, {Jiang}, {Zhang}, {Zhu}, {Shu}, {Huang}, {Zhang}, {Sheng}, \& {Lin}}]{WWJ2024}
{Wang}, Y., {Wang}, T., {Jiang}, N., {et~al.} 2024, \apj, 966, 136, \dodoi{10.3847/1538-4357/ad2ae4}

\bibitem[{{Wevers} {et~al.}(2023){Wevers}, {Coughlin}, {Pasham}, {Guolo}, {Sun}, {Wen}, {Jonker}, {Zabludoff}, {Malyali}, {Arcodia}, {Liu}, {Merloni}, {Rau}, {Grotova}, {Short}, \& {Cao}}]{Wevers+2023ApJ}
{Wevers}, T., {Coughlin}, E.~R., {Pasham}, D.~R., {et~al.} 2023, \apjl, 942, L33, \dodoi{10.3847/2041-8213/ac9f36}

\bibitem[{{Yao} {et~al.}(2023){Yao}, {Ravi}, {Gezari}, {van Velzen}, {Lu}, {Schulze}, {Somalwar}, {Kulkarni}, {Hammerstein}, {Nicholl}, {Graham}, {Perley}, {Cenko}, {Stein}, {Ricarte}, {Chadayammuri}, {Quataert}, {Bellm}, {Bloom}, {Dekany}, {Drake}, {Groom}, {Mahabal}, {Prince}, {Riddle}, {Rusholme}, {Sharma}, {Sollerman}, \& {Yan}}]{Yao+2023ApJ}
{Yao}, Y., {Ravi}, V., {Gezari}, S., {et~al.} 2023, \apjl, 955, L6, \dodoi{10.3847/2041-8213/acf216}

\bibitem[{{Zhong} {et~al.}(2023){Zhong}, {Hayasaki}, {Li}, {Berczik}, \& {Spurzem}}]{ZHL2023ApJ}
{Zhong}, S., {Hayasaki}, K., {Li}, S., {Berczik}, P., \& {Spurzem}, R. 2023, \apj, 959, 19, \dodoi{10.3847/1538-4357/ad0122}

\bibitem[{{Zhou} {et~al.}(2021){Zhou}, {Liu}, {Komossa}, {Cao}, {Ho}, {Chen}, \& {Li}}]{ZLK2021}
{Zhou}, Z.~Q., {Liu}, F.~K., {Komossa}, S., {et~al.} 2021, \apj, 907, 77, \dodoi{10.3847/1538-4357/abcccb}

\end{thebibliography}

\bibliographystyle{aasjournal}



\appendix

\section{The radiation efficiency of stream-disk collision}
\label{Appendix-eta_SD}

The accurate luminosity of stream-disk collision should be measured in hydrodynamic  simulation~\citep{Ryu+2023ApJ.957.12R,Huang+2024ApJ}, which is beyond the scope of this paper. In what follows, we make a simple estimation on the contribution of the stream-disk collision based on the radiation efficiency. The amount of specific orbital energy dissipated in the collision could be evaluated as $\Delta \epsilon = \epsilon_{i} - \epsilon_{f}$, where $\epsilon_{i}$ is the specific orbital energy of the fallback debris and $\epsilon_{f}$ is the specific orbital energy after the collision. \cite{Huang+2024ApJ} has shown that after a few $t_{\rm min}$, a significant part of the debris accumulated in the region between $r_{\rm p}$ and the self-intersection radius $r_{\rm SI}$, and orbits around the MBH with moderate eccentricity. For simplicity, we assume that after the stream-disk collision, the debris settles down into an eccentric orbit with pericentric radius $r_{\rm p}$ and apocentric radius $r_{\rm a}=r_{\rm SI}$. The semi-major axis of this orbit is given by $a = (r_{\rm p}+r_{\rm SI})/2$, and the specific orbital energy is $\epsilon_f = -GM_{\rm BH}/(r_{\rm p}+r_{\rm SI})$. We compute $r_{\rm SI}$ using equation 7 of \cite{Dai+2015ApJ}. Because $|\epsilon_i| \ll |\epsilon_f|$, we can approximate $\epsilon_i=0$ and get $\Delta \epsilon \simeq GM_{\rm BH}/(r_{\rm p}+r_{\rm SI})$. Assuming the dissipated energy is 100 percent converted to radiation, we get the radiation efficiency of stream-disk collision as $\eta_{\rm SD} = r_{\rm g}/(r_{\rm p}+r_{\rm SI})$, where $r_{\rm g}=GM_{\rm BH}/c^2$ is the gravitational radius. Note, if we adopt the scenario of \cite{Steinberg+Stone2024Natur}, $\eta_{\rm SD}$ should be larger than the values given in Table~\ref{table-params2}. Due to the efficient circularization in their simulation, the size of the gas disk should smaller than $ (r_{\rm p}+r_{\rm SI})/2$.

\end{document}